\documentclass[a4paper,12pt]{article}
\usepackage{jheppub} 

\title{\boldmath $\mathcal{N}=2$ dilaton-Weyl multiplets in 5D and 
Nishino-Rajpoot supergravity off-shell}
\author{Peter Sloane}
\affiliation{Departamento de Ciencias F\'isicas,
Universidad Andr\'es Bello,
Rep\'ublica 220, Santiago, Chile.}
\emailAdd{peter.sloane@unab.cl}
\abstract{
			We describe in detail the derivation of a 
superconformal 
off-shell formulation of the alternative $\mathcal{N}=2$, $d=5$ ungauged 
supergravity of Nishino and Rajpoot, coupled to $n$ Abelian vector multiplets, 
using a general dilaton-Weyl multiplet. We generalize the vector multiplet 
coupling available in the literature and show under which assumptions that the 
scalar manifold reduces to the known case of SO(1,1) $\times$ SO(1,n)/SO(n). As 
an application of the formalism we propose generalized vector multiplet coupled 
higher curvature terms, whose construction we sketch briefly. }

		
\begin{document}
\maketitle
\flushbottom
\section{Introduction.}

A conventional on-shell formulation of $\mathcal{N}=2$ supergravity in five 
dimensions was initially given in \cite{Cremmer:1980gs,Chamseddine:1980sp} and 
the U(1) gauged case was first described in \cite{D'Auria:1981kq}. In 
\cite{Gunaydin:1984pf,Gunaydin:1983rk,Gunaydin:1983bi,Gunaydin:1984ak} on-shell 
methods were used to treat the case of this supergravity coupled to vector 
multiplets. Hypermultiplet couplings and gaugings were considered in 
\cite{Sierra:1985ax,Lukas:1998yy,Lukas:1998tt} and tensor multiplet matter in 
\cite{Gunaydin:1999zx,Gunaydin:2000xk} along with gaugings of isometries of a 
subgroup of the isometry group of the scalar manifold.
The theory can also be obtained from compactification of M-theory on a 
Calabi-Yau threefold $CY_{3}$ \cite{Cadavid:1995bk,Antoniadis:1995vz}. The 
resulting Lagrangian depends on topological data of the compactification 
manifold, namely the Calabi-Yau intersection numbers.  

This formulation of supergravity doesn't include the N-S two-form $B_{\mu\nu}$ 
and dilaton explicitly and in order to investigate effective descriptions of 
string theory it became important to include the dilaton and antisymmetric 
fields, so off-shell formulations 
\cite{Howe:1981ev,Howe:1981nz,Zucker:1999ej,Zucker:1999fn,Kugo:2000hn,Kugo:2000af} were explored to 
facilitate the construction of matter coupled supergravities, although these 
theories lack a manifest $\sigma$-model structure for the scalars before 
eliminating the auxiliary fields \cite{Cremmer:1982en}. 
In \cite{Nishino:2000cz}, Nishino and Rajpoot proposed an alternative on-shell 
formulation of $\mathcal{N}=2$ $d=5$ supergravity starting from a supergravity 
multiplet with a larger field content which contains the N-S antisymmetric 
field 
$B_{\mu\nu}$ and a dilaton $\sigma$. This multiplet's vielbein ${e_{\mu}}^{m}$, 
gravitini ${\psi_{\mu}}^{\mathbf{i}}$ and graviphoton $A_{\mu}$ coincide with 
the conventional fields and in addition to the two-form and dilaton, there is a 
dilatino $\chi^{\mathbf{i}}$, giving rise to $12+12$ on-shell degrees of 
freedom. Vector and hypermultiplets \cite{Nishino:2001ji} have been coupled to 
this supergravity theory, with a structure of the couplings similar to that of 
$\mathcal{N}=1$ $d=9$ supergravity \cite{Gates:1984kr}. 
A priori, both formulations are rather similar if one dualizes the 
antisymmetric 
tensor $B_{\mu\nu}$ into a vector field $B_{\mu}$. However, after coupling to 
vector multiplets, the resulting $\sigma$-model structure is different. In 
fact, 
it was shown in \cite{Bergshoeff:2001hc} that the dilaton-Weyl multiplet can be 
obtained by coupling the standard multiplet to an improved vector multiplet.

The matter couplings of $\mathcal{N}=2$ d=5 supergravity were studied 
extensively in \cite{Ceresole:2000jd} from a superspace perspective, and 
further 
work using the superconformal formulation 
\cite{Bergshoeff:2001hc,Fujita:2001kv} 
allowed the construction of superconformal multiplets and their corresponding 
actions \cite{Bergshoeff:2001hc,Fujita:2001kv,Kugo:2000hn, 
Kugo:2000af,Bergshoeff:2002qk}, leading to quite general $d=5$ matter couplings 
in the superconformal formulation \cite{Bergshoeff:2004kh}. The resulting 
theories preserve eight supersymmetries\footnote{We use the terminology 
$\mathcal{N}=2$ due to the fact we use symplectic Majorana spinors, which are 
an 
SU(2) doublet of complex spinors obeying the symplectic Majorana condition, 
$\psi^\mathbf{i}=\epsilon^{\mathbf{ij}}(\psi^{\mathbf{j}})^{c}$ where $c$ 
denotes charge conjugation. In the literature sometimes the notation 
$\mathcal{N}=1$ is used in the case that the theory is presented in terms of 
Dirac spinors. Of course these two descriptions both have 8 real components of 
the supercharges.} \cite{VanProeyen:2001wr} and can be studied at depth with 
the 
tools of special geometry 
\cite{deWit:1983rz,deWit:1984pk,Gates:1983py,Sierra:1983cc}, the condition for 
which arises in the off-shell theory as a constraint coming from a scalar 
Lagrange multiplier auxiliary field of the standard-Weyl multiplet. The 
advantage of the off-shell formulation is that we may find higher derivative 
densities, which are important from a string theory perspective, without 
changing the supersymmetry transformations, and therefore inducing corrections 
to our original action, an iterative process that may never terminate. The 
higher derivative densities that are supersymmetric completions of the square 
of 
the Ricci scalar and the square of the Weyl tensor have been produced in the 
background of the standard-Weyl superconformal gravitational multiplet in 
\cite{Hanaki:2006pj,Ozkan:2013nwa}. 

In \cite{Bergshoeff:2001hc} dilaton-Weyl multiplets were introduced including 
the two form, the dilaton and the dilatino, whilst in \cite{Fujita:2001kv} 
dilaton-Weyl multiplets incorporating more than one vector multiplet were 
introduced. In 
\cite{Kuzenko:2006mv,Kuzenko:2007cj,Kuzenko:2007hu,Kuzenko:2008wr,
Kuzenko:2013rna,Kuzenko:2014eqa} an off-shell superspace formulation of the 
superconformal theory has been developed, which should lead to the most general 
couplings, and indeed the dilaton-Weyl multiplet was considered in these works. 
We find it useful to add to the literature an explicit derivation of the N-R 
supergravity from the off-shell formulation by means of gauge fixing and field 
redefinitions, complimenting the work done in \cite{Fujita:2001kv}. We shall 
discuss in detail the vector multiplet couplings of this theory. We shall also 
discuss simple generalizations of two of the higher derivative densities 
\cite{Hanaki:2006pj,Ozkan:2013nwa} found in the literature. 

This paper is organized as follows. In section \ref{Pure} we discuss the 
derivation of the minimal N-R supergravity and in section 
\ref{off-shellvectorcoupled} we couple to Abelian vector multiplets and 
relegate 
to appendix \ref{explictredef} the explicit constant field redefinitions needed 
to arrive at the conventions of \cite{Nishino:2000cz,Nishino:2001ji}. In 
section 
\ref{Higher} we generalize the known higher derivative densities to the 
extended 
dilaton-Weyl multiplets that we describe in appendix \ref{gendilatonweyl}, in 
which we make use of a composition of a vector multiplet in terms of a linear 
multiplet \cite{Coomans:2012cf} that we give in appendix \ref{vlcomposition}. 
We 
conclude in section \ref{Conclusions}.

\acknowledgments{The work of the author is supported by FONDECYT Postdoctorado 
Project number 3130541. The author would like to thank Linda Uruchurtu, Jorge 
Bellor\'in and Hitoshi Nishino for useful correspondence and discussions, and 
Per Sundell and Rodrigo Olea for encouragement.}

\section{Pure N-R supergravity from the off-shell superconformal 
formalism.}\label{Pure}

In this section we give the details of the construction of the N-R supergravity 
\cite{Nishino:2000cz,Nishino:2001ji} from the off-shell formalism based on the 
superconformal dilaton-Weyl multiplet described in \cite{Coomans:2012cf}. We 
also describe an alternative procedure put forward in \cite{Fujita:2001kv}. 
To couple the theory of \cite{Coomans:2012cf} to vector multiplets one may use 
the results of \cite{Ozkan:2013nwa}, however following the procedure of 
\cite{Fujita:2001kv} we will be led to introduce a larger generalized 
dilaton-Weyl multiplet, which includes an arbitrary number of vector 
multiplets. 
It is instructive to consider the case of the pure N-R supergravity first, and 
then the coupling to vector multiplets separately.

The pure N-R supergravity can be constructed straightforwardly using exactly 
the 
results of \cite{Coomans:2012cf}, whose conventions we will follow, which are 
described in detail in \cite{Bergshoeff:2001hc}. However we shall construct it 
in a slightly different way that was suggested in \cite{Fujita:2001kv}, as we 
will emphasize below. The two derivative theory is constructed by combining a 
vector multiplet action and a compensating linear multiplet action, obtained in 
the background of a Weyl multiplet. We suppress the spinor indices in bilinears 
using the NW-SE convention and we raise and lower the SU(2) indices using the 
totally antisymmetric tensor $\epsilon_{\mathbf{ij}}$ where 
$\epsilon_{\mathbf{12}}=\epsilon^{\mathbf{12}}=1$, e.g. $\bar{\psi}_\mu 
\psi_\nu=\bar{\psi}_\mu^{\mathbf{i}}\psi_{\mathbf{i}\nu}=\bar{\psi}_\mu^{\mathbf
{i}}\psi^{\mathbf{j}}_{\nu}\epsilon_{\mathbf{ji}}$. We will frequently use the 
notation that for two p-forms $\alpha,\beta$, we define $\alpha\cdot 
\beta=\alpha_{\mu_1\cdots\mu_p}\beta^{\mu_1\cdots\mu_p}$, and 
$\alpha^2=\alpha\cdot\alpha$.

There are two types of Weyl multiplet, the so called standard-Weyl multiplet 
and 
the dilaton-Weyl multiplet. The standard-Weyl multiplet consists of the 
vielbien 
$e_\mu^m$, gravitino $\psi_\mu^{\mathbf{i}}$, an auxiliary two form $T_{mn}$,
an auxiliary scalar $D$, an auxiliary fermion
$\chi^{\mathbf{i}}$, an auxiliary SU(2) 
triplet of vectors $V^{\mathbf{ij}}_\mu$ with 
$V^{\mathbf{ij}}_\mu=V^{\mathbf{ji}}_{\mu}$ and a gauge field for local 
dilatations, $b_{\mu}$.
These transform under supersymmetry with parameter $\epsilon^\mathbf{i}$ and 
special supersymmetry with parameter $\eta^\mathbf{i}$ as
\begin{align}
 \delta e_\mu^m &=\tfrac{1}{2}\bar{\epsilon}\gamma^m\psi_\mu \ ,\nonumber\\
 \delta \psi_\mu^{\mathbf{i}} &= (\nabla_\mu 
+\tfrac{1}{2}b_\mu)\epsilon^{\mathbf{i}} 
-V_{\mu}^{\mathbf{ij}}\epsilon_{\mathbf{j}} 
+i\gamma_{mn}{T}^{mn}\gamma_\mu\epsilon^{\mathbf{i}} - 
i\gamma_\mu\eta^{\mathbf{i}}\ ,\nonumber\\
 \delta V^{\mathbf{ij}}_\mu &= -\tfrac{3i}{2}\bar{\epsilon}^{\mathbf{(i}} 
\underline{\phi}^{\mathbf{j)}}_\mu +4\bar{\epsilon}^{\mathbf{(i}}\gamma_\mu 
\chi^{\mathbf{j)}} + 
i\bar{\epsilon}^{\mathbf{(i}}\gamma_{mn}{T}^{mn}\psi^{\mathbf{j)}}_\mu 
+\tfrac{3i}{2}\bar{\eta}^{\mathbf{(i}}\psi^{\mathbf{j)}}_\mu \ ,\nonumber\\
 \delta T_{mn} &= \tfrac{i}{2}\bar{\epsilon}\gamma_{mn}\chi - \tfrac{3i}{32} 
\bar{\epsilon}\hat{\underline{R}}_{mn}(Q) \ ,\nonumber\\
 \delta \chi^{\mathbf{i}} &= \tfrac{1}{4}D\epsilon^{\mathbf{i}} - 
\tfrac{1}{64}\gamma^{mn}\hat{\underline{R}}^{\mathbf{ij}}_{mn}(V)\epsilon_{
\mathbf{j}} + 
\tfrac{i}{8}\gamma^{mn}\gamma_{p}\mathcal{D}^p{T}_{mn}\epsilon^{\mathbf{i}} 
\nonumber\\
 &- \tfrac{i}{8}\gamma^m\mathcal{D}^{n}T_{mn}\epsilon^{\mathbf{i}} 
-\tfrac{1}{4}\gamma^{mnpq}T_{mn}T_{pq}\epsilon^{\mathbf{i}} 
+\tfrac{1}{6}T^2\epsilon^{\mathbf{i}} + 
\tfrac{1}{4}\gamma_{mn}T^{mn}\eta^{\mathbf{i}}\ ,
 \nonumber\\
 \delta D &= \bar{\epsilon}\gamma^m\mathcal{D}_m\chi 
-\tfrac{5i}{3}\bar{\epsilon}\gamma_{mn}T^{mn}\chi -i\bar{\eta}\chi \ 
,\nonumber\\
 \delta b_\mu &= \tfrac{i}{2}\bar{\epsilon}\underline{\phi}_\mu - 
2\bar{\epsilon}\gamma_\mu\chi + \tfrac{i}{2}\bar{\eta}\psi_\mu \ ,
\end{align}
where the the spin covariant derivative is defined by
\begin{equation}
 \nabla_\mu\epsilon^\mathbf{i}=(\partial_\mu 
+\tfrac{1}{4}{{\omega}_{\mu}}^{mn}\gamma_{mn})\epsilon^{\mathbf{i}} \ ,
\end{equation}
and we have underlined the composite fields apart from the spin connection. 
Explicit expressions for the composite fields are
\begin{align}
{{\omega}}_\mu^{mn}&= 2e^{\nu[m}\partial_{[\mu} e^{n]}_{\nu]} - 
e^{\nu[m}e^{n]}\sigma e_{\mu p}\partial_\nu e_\sigma^p + 2e_{\mu}^{[m}b^{n]} 
-\tfrac{1}{2}\bar{\psi}^{[n}\gamma^{m]}\psi_\mu 
-\tfrac{1}{4}\bar{\psi}^n\gamma_\mu\psi^m \ ,\nonumber\\
\underline{\phi}^\mathbf{i}_\mu &=\tfrac{i}{3}\gamma^m\underline{\hat{R'}}_{\mu 
m}^{\mathbf{i}}(Q)-\tfrac{i}{24}\gamma_\mu \gamma^{mn} 
{\hat{\underline{R}'}}_{mn}^{\mathbf{i}}(Q)\ , \nonumber\\
\underline{\hat{R'}}_{\mu \nu}^{\mathbf{i}}(Q) &= 
2\nabla_{[\mu}\psi^{\mathbf{i}}_{\nu]} + b_{[\mu}\psi^{\mathbf{i}}_{\nu]} - 
2V^{\mathbf{ij}}_{[\mu}\psi_{\nu]\mathbf{j}} + 
2i\gamma_{mn}T^{mn}\gamma_{[\mu}\psi^{\mathbf{i}}_{\nu]}\ ,\nonumber\\
\underline{\hat{R}}_{\mu \nu}^{\mathbf{i}}(Q) &=\underline{\hat{R'}}_{\mu 
\nu}^{\mathbf{i}}(Q) -2i\gamma_{[\mu}\underline{\phi}_{\nu]}^{\mathbf{i}} \ 
,\nonumber\\
\underline{\hat{R}}^{\mathbf{ij}}_{\mu\nu}(V) &= 
2\partial_{[\mu}V^{\mathbf{ij}}_{\nu]} 
-2V_{[\mu}^{\mathbf{k(i}}V_{\nu]\mathbf{k}}^{\mathbf{j)}}-3i{\bar{\underline{
\phi}}}_{[\mu}^{\mathbf{(i}}\psi^{\mathbf{j)}}_{\nu]} 
-8\bar{\psi}^{\mathbf{(i}}_{[\mu}\gamma_{\nu]}\chi^{\mathbf{j)}} 
-i\bar{\psi}^{\mathbf{(i}}_{[\mu} (\gamma\cdot T)\psi^{\mathbf{j)}}_{\nu]}\ , 
\nonumber\\
{\underline{R'}(M)_{\mu\nu}}^{mn} &=2 \partial_{[\mu}{{\omega}_{\nu]}}^{mn} 
+2{{\omega}_{[\mu}}^{mp}{{\omega}_{\nu]p}}^n 
+i\bar{\psi}_{[\mu}\gamma^{mn}\psi_{\nu]} + 
i\bar{\psi}_{[\mu}\gamma^{[m}(\gamma\cdot T)\gamma^{n]}\psi_{\nu]}\nonumber\\
& 
+\bar{\psi}_{[\mu}\gamma^{[m}{\underline{R}_{\nu]}}^{n]}(Q)+\tfrac{1}{2}\bar{
\psi}_{[\mu}\gamma_{\nu]}\underline{R}^{mn}(Q) 
-8\bar{\psi}_{[\mu}{e_{\nu]}}^{[m}\gamma^{n]}\chi  + 
i\bar{\underline{\phi}}_{[\mu}\gamma^{mn}\psi_{\nu]}\ , \nonumber\\
\underline{f}_m^m &=-\tfrac{1}{16}\underline{\mathcal{R}} \ , \qquad 
\mathcal{\mathcal{R}}={\underline{R'}(M)_{\mu\nu}}^{\mu\nu}\ ,  
\label{compositesuperconformalfields}
\end{align}
where the relevant superconformal derivatives are given by
\begin{align}
 \mathcal{D}_\mu\chi^{\mathbf{i}} &= (\nabla_\mu -\tfrac{3}{2}b_\mu) 
\chi^{\mathbf{i}} - V^{\mathbf{ij}}_\mu \chi_{\mathbf{j}} 
-\tfrac{1}{4}D\psi_\mu^{\mathbf{i}} 
+\tfrac{1}{64}\gamma^{mn}\hat{R}^{\mathbf{ij}}_{mn}(V) \psi_{\mu \mathbf{j}} 
-\tfrac{i}{8}\gamma^{mn}\gamma^p (\mathcal{D}_p T_{mn}) \psi^{\mathbf{i}}_\mu 
\nonumber\\
& +\tfrac{i}{8}\gamma^m(\mathcal{D}^n T_{mn})\psi^{\mathbf{i}}_\mu + 
\tfrac{1}{4}\gamma^{mnpq}T_{mn}T_{pq}\psi^{\mathbf{i}}_\mu 
-\tfrac{1}{6}T^2\psi^{\mathbf{i}}_\mu 
-\tfrac{1}{4}\gamma_{mn}T^{mn}\underline{\phi}^{\mathbf{i}}_\mu \ ,
\end{align}
and
\begin{equation}
\mathcal{D}_\mu T_{mn} = (\nabla_\mu -b_\mu) T_{mn} -\tfrac{i}{2}\bar{\psi}_\mu 
\gamma_{mn}\chi +\tfrac{3i}{32}\bar{\psi}_\mu\hat{R}_{mn}(Q)\ .
\end{equation}

The superconformal linear multiplet is formed from an SU(2) triplet 
$L^{\mathbf{ij}}=L^\mathbf{ji}$, a constrained vector $E_m$, a scalar $N$  and 
a 
fermion $\varphi^{\mathbf{i}}$ which transform, in the background of the 
standard-Weyl multiplet, as
\begin{align}
 \delta L^{\mathbf{ij}} &=i\bar{\epsilon}^{\mathbf{(i}}\varphi^{\mathbf{j)}} \ 
,\nonumber\\
 \delta 
\varphi^{\mathbf{i}}&=-\tfrac{i}{2}\gamma^m\mathcal{D}_mL^{\mathbf{ij}}\epsilon_
{\mathbf{j}} 
-\tfrac{i}{2}\gamma^mE_m\epsilon^{\mathbf{i}}+\tfrac{N}{2}\epsilon^{i}-\gamma_{
mn}T^{mn}L^{\mathbf{ij}}\epsilon_{\mathbf{j}} 
+3L^{\mathbf{ij}}\eta_{\mathbf{j}}\ ,\nonumber\\
 \delta E_m &= -\tfrac{i}{2}\bar{\epsilon}\gamma_{mn}\mathcal{D}^n\varphi 
-2\bar{\epsilon}\gamma^n\varphi T_{nm} - 2\bar{\eta}\gamma_{m}\varphi \ , 
\nonumber\\
 \delta N &= \tfrac{1}{2}\bar{\epsilon}\gamma^m\mathcal{D}_{m}\varphi 
+\tfrac{3i}{2}\bar{\epsilon}\gamma_{mn}T^{mn}\varphi + 
4i\bar{\epsilon}^{\mathbf{i}}\chi^{\mathbf{i}}L_{\mathbf{ij}} 
+\tfrac{3i}{2}\bar{\eta}\varphi \ ,
 \end{align}
 where
 \begin{align}
  \mathcal{D}_\mu L^{\mathbf{ij}}&=(\partial_\mu -3b_\mu)L^{\mathbf{ij}} + 
2V^{\mathbf{(i}}_{\mu\phantom{(i}\mathbf{k}}L^{\mathbf{j)k}} - 
i\bar{\psi}_\mu^{\mathbf{(i}}\varphi^{\mathbf{j)}} \ ,\nonumber\\  
\mathcal{D}_\mu\varphi^{\mathbf{i}}&=(\nabla_\mu-\tfrac{7}{2}b_\mu)\varphi^{
\mathbf{i}} -V_\mu^{\mathbf{ij}}\varphi_{\mathbf{j}} - 
\tfrac{i}{2}\gamma^m\mathcal{D}_mL^{\mathbf{ij}}\psi_{\mu \mathbf{j}} + 
\tfrac{i}{2}\gamma^mE_m\psi^{\mathbf{i}}_\mu - 
\tfrac{N}{2}\psi^{\mathbf{i}}_\mu 
\nonumber\\
  &+ \gamma_{mn}T^{mn}L^{\mathbf{ij}}\psi_{\mu\mathbf{j}} 
-3L^{\mathbf{ij}}\phi_{\mu\mathbf{j}} \ ,\nonumber\\
  \mathcal{D}_\mu E_m &=(\nabla_\mu -4b_\mu )E_m 
+\tfrac{i}{2}\bar{\psi}_\mu\gamma_{mn}\mathcal{D}^n\varphi + 
2\bar{\psi}_\mu\gamma^n\varphi T_{nm} 
+2\bar{\underline{\phi}}_\mu\gamma_m\varphi \ .
 \end{align}

 The constraint on the vector $E^m$, which reads $\mathcal{D}^mE_m=0$ can be 
solved by the introduction of a three form $E_{\mu\nu\rho}$ such that
 \begin{equation}  
E^m=-\tfrac{1}{12}e_\mu^me^{-1}\epsilon^{\mu\nu\rho\sigma\lambda}\mathcal{D}_{
\nu}E_{\rho\sigma\lambda} \ ,
 \end{equation}
 and it is useful to define the two form 
$E_{\mu\nu\rho}=e\epsilon_{\mu\nu\rho\sigma\lambda}E^{\sigma\lambda}$, so that 
we have $E^m=e^m_\mu \mathcal{D}_\nu E^{\mu\nu}$.

 The vector multiplet is
 formed from an SU(2) triplet of scalars $Y^{\mathbf{ij}}$, the gauge field 
$A_\mu$, a gaugino $\lambda^{\mathbf{i}}$ and a scalar $\rho$. These transform 
under the supersymmetries in the background of the standard-Weyl multiplet as
 \begin{align}
  \delta A_\mu &= -\tfrac{i}{2}\rho \bar{\epsilon}\psi_{\mu} + 
\tfrac{1}{2}\bar{\epsilon}\gamma_\mu\lambda \ ,\nonumber\\
  \delta Y^{\mathbf{ij}} &= 
-\tfrac{1}{2}\bar{\epsilon}^{\mathbf{(i}}\gamma^m\mathcal{D}_m\lambda^{\mathbf{
j)}} + \tfrac{i}{2}\bar{\epsilon}^{\mathbf{(i}}(\gamma\cdot 
T)\lambda^{\mathbf{j)}} - 4i\rho\bar{\epsilon}^{(\mathbf{i}}\chi^{\mathbf{j)}} + 
\tfrac{i}{2}\bar{\eta}^{\mathbf{(i}}\lambda^{\mathbf{j)}} \ , \nonumber\\
  \delta \lambda^{\mathbf{i}} &= -\tfrac{1}{4}\gamma_{mn}\hat{F}^{mn} 
\epsilon^{\mathbf{i}} - 
\tfrac{i}{2}\gamma^m(\mathcal{D}_m\rho)\epsilon^{\mathbf{i}} +\rho 
\gamma_{mn}T^{mn}\epsilon^{\mathbf{i}} - Y^{\mathbf{ij}}\epsilon_{\mathbf{j}} 
+\rho\eta^{\mathbf{i}} \ ,\nonumber\\
  \delta \rho &=\tfrac{i}{2}\bar{\epsilon}\lambda \ , 
 \end{align}
 where
 \begin{align}
  \mathcal{D}_\mu \rho &= (\partial_\mu -b_\mu)\rho 
-\tfrac{i}{2}\bar{\psi}_\mu\lambda \ , \nonumber\\
  \mathcal{D}_\mu \lambda^{\mathbf{i}} &= (\nabla_\mu -\tfrac{3}{2}b_\mu) 
\lambda^{\mathbf{i}} -V_{\mu}^{\mathbf{ij}}\lambda_{\mathbf{j}} + 
\tfrac{1}{4}\gamma_{mn}\hat{F}^{mn}\psi^{\mathbf{i}}_\mu + 
\tfrac{i}{2}\gamma^m(\mathcal{D}_m\rho)\psi^{\mathbf{i}}_\mu \nonumber\\&
  +Y^{\mathbf{ij}}\psi_{\mu\mathbf{j}} - \rho\gamma_{mn} 
T^{mn}\psi^{\mathbf{i}}_\mu -\rho\underline{\phi}^{\mathbf{i}}_\mu \ 
,\nonumber\\
  \hat{F}_{\mu\nu}&=F_{\mu\nu} -\bar{\psi}_{[\mu}\gamma_{\nu]}\lambda 
+\tfrac{i}{2}\rho\bar{\psi}_{[\mu}\psi_{\nu]} \ ,
 \end{align}
 and where $F=dA$.

 A superconformally invariant density formula constructed from a vector 
multiplet and a linear multiplet is given by
 \begin{align}
  e^{-1}\mathcal{L}_{VL} &= Y^{\mathbf{ij}}L_{\mathbf{ij}} + 
i\bar{\lambda}\varphi 
-\tfrac{1}{2}\bar{\psi}_m^{\mathbf{i}}\gamma^m\lambda^{\mathbf{j}}L_{\mathbf{ij}
} + C_mP^m \nonumber\\
  &+ \rho\left(N +\tfrac{1}{2}\bar{\psi}_m\gamma^m\varphi 
+\tfrac{i}{4}\bar{\psi}_m^{\mathbf{i}}\gamma^{mn}\psi_n^{\mathbf{j}}L_{\mathbf{
ij}} \right) \ , \label{actionformula}
 \end{align}
where $P^m$ is the bosonic part of the supercovariant $E^m$
\begin{equation}
 P^m=E^m +\tfrac{i}{2}\bar{\psi}_n\gamma^{nm}\varphi 
+\tfrac{1}{4}\bar{\psi}_n^{\mathbf{i}}\gamma^{mnp}\psi^{\mathbf{j}}_p 
L_{\mathbf{ij}} \ .
\end{equation}

 In order to describe vector-vector couplings one can compose the linear 
multiplet appearing in the above action from a vector multiplet 
and to describe linear-linear couplings one can compose the vector multiplet 
appearing in the action from a linear multiplet.  
The composition of the vector multiplet from the linear multiplet is given in 
detail in \cite{Coomans:2012cf,Ozkan:2013nwa} and we list the bosonic parts in 
appendix \ref{compositevectoroflinear}. As noted in \cite{Fujita:2001kv}, where 
only the scalar composition was given, this embedding leads to fairly long 
expressions when including the fermions. We will be interested in the bosonic 
part of the resulting action which reads  
 \begin{align}
  e^{-1}\mathcal{L}_L &= L^{-1}L_{\mathbf{ij}}\Box L^{\mathbf{ij}} - 
L^{\mathbf{ij}}\mathcal{D}_{\mu}L_{\mathbf{k(i}}\mathcal{D}^\mu 
L_{\mathbf{j)m}}L^{\mathbf{km}} L^{-3} -N^2 L^{-1} \nonumber\\
  &- P_\mu P^\mu L^{-1} +\tfrac{8}{3}L {T}^2 +4{D}L  
-\tfrac{1}{2}L^{-3}P^{\mu\nu}L^{\mathbf{l}}_{\mathbf{k}}\partial_\mu 
L^{\mathbf{kp}}\partial_\nu L_{\mathbf{pl}} \nonumber\\
  &+ 2P^{\mu\nu}\partial_\mu(L^{-1} P_\nu 
+V_\nu^{\mathbf{ij}}L_{\mathbf{ij}}L^{-1}) \ ,  \label{LAGLstandard}
 \end{align}
 where $L^2=L_{\mathbf{ij}}L^{\mathbf{ij}}$, $P^{\mu\nu}$ is the bosonic part 
of 
$E^{\mu\nu}$ and
 \begin{align}
  L_{\mathbf{ij}}\Box L^{\mathbf{ij}} &= L_{\mathbf{ij}}(\partial^m -4b^m 
+{\omega_n}^{nm})\mathcal{D}_mL^{\mathbf{ij}} 
+2L_{\mathbf{ij}}V_{n\phantom{\mathbf{i}}\mathbf{k}}^{\mathbf{i}}\mathcal{D}^nL^
{\mathbf{jk}} \nonumber\\
  &+6L^2\underline{f}^m_m 
  -iL_{\mathbf{ij}}\bar{\psi}^{m\mathbf{i}}\mathcal{D}_m\varphi^{\mathbf{j}} 
-6L^2\bar{\psi}^m\gamma_m{\chi} \nonumber\\  
&-L_{\mathbf{ij}}\bar{\varphi}^{\mathbf{i}}\gamma_{mn}{T}^{mn}\gamma^p\psi_p^{
\mathbf{j}} 
+L_{\mathbf{ij}}\bar{\varphi}^{\mathbf{i}}\gamma^m{\underline{\phi}}_m^{\mathbf{
j}} 
  \ .
 \end{align}
The composition of the linear multiplet in terms of a single vector multiplet 
is 
well known \cite{Kugo:2000af,Bergshoeff:2001hc,Fujita:2001kv}, which we take 
from (A.1) of \cite{Ozkan:2013nwa},\footnote{We have corrected a typo of a 
missing factor of $\rho$ in the last term of the first line of the expression 
for $N(\mathbf{V})$ and a missing factor of $i$ in the penultimate term in the 
expression for $\varphi^{\mathbf{i}}(\mathbf{V})$.} and reads
\begin{align}
 L_{\mathbf{ij}}(\mathbf{V}) &= 2\rho Y_{\mathbf{ij}} 
-\tfrac{i}{2}\bar{\lambda}_{\mathbf{i}}\lambda_{\mathbf{j}} \ , \nonumber\\
 \varphi_{\mathbf{i}}(\mathbf{V}) &= 
i\rho\gamma^m\mathcal{D}_m\lambda_{\mathbf{i}} + 
2\rho\gamma_{mn}T^{mn}\lambda_{\mathbf{i}} -8\rho^2\chi_{\mathbf{i}} - 
\tfrac{1}{4}\gamma^{mn}\hat{F}_{mn}\lambda_{\mathbf{i}} + 
\tfrac{i}{2}\gamma^m(\mathcal{D}_m\rho)\lambda_{\mathbf{i}} 
-Y_{\mathbf{ij}}\lambda^{\mathbf{j}}\ , \nonumber\\
 E^m(\mathbf{V}) &= \mathcal{D}_n\left(-\rho \hat{F}^{mn} +8\rho^2 T^{mn} 
-\tfrac{i}{4}\bar{\lambda}\gamma_{mn}\lambda \right) 
-\tfrac{1}{8}\epsilon^{mnpqr}\hat{F}_{np}\hat{F}_{qr} \ ,\nonumber\\
 N(\mathbf{V}) &= \rho \Box \rho + \tfrac{1}{2}(\mathcal{D}_m \rho) 
(\mathcal{D}^m\rho) - \tfrac{1}{4}\hat{F}_{mn}\hat{F}^{mn} 
+Y^{\mathbf{ij}}Y_{\mathbf{ij}} +8\rho\hat{F}_{mn}T^{mn} \nonumber\\
 &- 4\rho^2\left(D+\tfrac{26}{3}T^2\right) 
-\tfrac{1}{2}\bar{\lambda}\gamma^m\mathcal{D}_m\lambda 
+i\bar{\lambda}\gamma_{mn}T^{mn}\lambda + 16i\rho\bar{\chi}\lambda \ .
\end{align}

With this at hand we can now write down an action by taking the Lagrangian 
$\mathcal{L}_L -3\mathcal{L}_{\mathbf{V}}$, where
$\mathcal{L}_{\mathbf{V}}$ can be formed in two ways: by taking another copy of 
the same vector multiplet 
$\mathbf{V}=(\rho,A_\mu,\lambda^{\mathbf{i}},Y^{\mathbf{ij}})$, or by 
considering a second vector multiplet.
Let us first consider using the same vector multiplet that we have embedded in 
the linear multiplet as done in \cite{Coomans:2012cf}.
We obtain for the bosonic part of the vector multiplet density
\begin{align}
 e^{-1}\mathcal{L}_V &= -\tfrac{1}{4}\rho F^2 +\tfrac{1}{3}\rho^2\Box \rho + 
\tfrac{\rho}{6}(\mathcal{D}\rho)^2 + \rho Y^{\mathbf{ij}}Y_{\mathbf{ij}} 
\nonumber\\
 &-\tfrac{4}{3}\rho^3\left(D+\tfrac{26}{3}T^2\right) + 
4\rho^2F_{\mu\nu}T^{\mu\nu}-\tfrac{e^{-1}}{24}\epsilon^{\mu\nu\rho\sigma\lambda}
A_\mu F_{\nu\rho} F_{\sigma\lambda} \ ,
\end{align}
where
\begin{align}
 \Box\rho &= (\nabla^m -2b^m)\mathcal{D}_m\rho - 
\tfrac{i}{2}\bar{\psi}_m\mathcal{D}^m\lambda 
-2\rho\bar{\psi}_m\gamma^m\underline{\chi} \nonumber\\&
 +\tfrac{1}{2}\bar{\psi}_m\gamma^m\gamma_{np}\underline{T}^{np}\lambda 
+\tfrac{1}{2}\bar{\underline{\phi}}^m\gamma_m\lambda + 2\rho\underline{f}_m^m  
.\nonumber
\end{align}
It turns out that the equations of motion for the vector multiplet fields 
imply\footnote{For the details see \cite{Coomans:2012cf} where the relevant 
fermionic terms in the action are given.}
 \begin{align}
\underline{T}^{mn} &= \tfrac{\rho^{-2}}{8}(\rho\hat{F}^{mn} + 
\tfrac{1}{6}\epsilon^{mnpqr}\hat{H}_{pqr} + 
\tfrac{i}{4}\bar{\lambda}\gamma^{mn}\lambda) \ ,\nonumber\\
\underline{\chi}^{\mathbf{i}}&=\tfrac{i}{8}\rho^{-1}\gamma^{m}\mathcal{D}
_m\lambda^{\mathbf{i}} 
+\tfrac{i}{16}\rho^{-2}\gamma^{m}(\mathcal{D}_m\rho)\lambda^{\mathbf{i}} - 
\tfrac{\rho^{-2}}{32}\gamma_{mn}\hat{F}^{mn}\lambda^{\mathbf{i}} \nonumber\\
&+ \tfrac{\rho^{-1}}{4}\gamma_{mn}\underline{T}^{mn}\lambda^{\mathbf{i}} 
+\tfrac{i\rho^{-1}}{32}\lambda_{\mathbf{j}}\bar{\lambda}^{\mathbf{i}}\lambda^{
\mathbf{j}}\ , \nonumber\\
\underline{D} &= \tfrac{\rho^{-1}}{4}\Box\rho 
+\tfrac{\rho^{-2}}{8}(\mathcal{D}\rho)^2 - \tfrac{\rho^{-2}}{16}\hat{F}^2 
-\tfrac{\rho^{-2}}{8}\bar{\lambda}\gamma^m\mathcal{D}_m\lambda - 
\tfrac{\rho^{-4}}{64}\bar{\lambda}^{\mathbf{i}}\lambda^{\mathbf{j}}\bar{\lambda}
_{\mathbf{i}}\lambda_{\mathbf{j}} -4i\rho^{-1}\lambda\underline{\chi} 
\nonumber\\
&+\left(2\rho^{-1}\hat{F}_{mn} -\tfrac{26}{3}\underline{T}_{mn}\ 
+\tfrac{i\rho^{-2}}{4}\bar{\lambda}\gamma_{mn}\lambda \right)\underline{T}^{mn} 
\ ,\nonumber\\ 
\underline{Y}^{\mathbf{ij}}&=\tfrac{i}{4}\rho^{-1}\bar{\lambda}^{\mathbf{i}}
\lambda^{\mathbf{j}}\ ,
 \end{align}
 where 
 \begin{eqnarray}
\hat{H}_{\mu\nu\rho} &=& H_{\mu\nu\rho} - 
\tfrac{3}{4}\rho^2\bar{\psi}_{[\mu}\gamma_{\nu}\psi_{\rho]} - 
\tfrac{3i}{2}\rho\bar{\psi}_{[\mu}\gamma_{\nu\rho]}\lambda \ ,\nonumber\\
H_{\mu\nu\rho} &=& 3\partial_{[\mu} B_{\nu\rho]} 
+\tfrac{3}{2}A_{[\mu}F_{\nu\rho]} \ , 
 \end{eqnarray}
and for $H$ to be gauge invariant we need that $B$ transforms under gauge 
transformations as
\begin{equation}
\delta B_{\mu\nu} = 2\partial_{[\mu}\Lambda_{\nu]} - \tfrac{1}{2}\Lambda 
F_{\mu\nu} \ .
\end{equation}
 Now we note that the equation of motion for $D$ is given by
 \begin{equation}
  L=\rho^3 \ .
 \end{equation}
This must be implemented as a constraint if one is to use the above solutions 
of 
the equations of motion in the action, and obtain an equivalent theory. However 
the gauge fixing performed in \cite{Coomans:2012cf,Ozkan:2013nwa} demands that 
$L$ be constant,
\begin{equation}
 L_{\mathbf{ij}}=\pm\tfrac{1}{\sqrt2}\delta_{\mathbf{ij}} \qquad b_\mu=0 \qquad 
\lambda=0 \ .\label{gaugefixing}
\end{equation}
So the action given in \cite{Coomans:2012cf} should be supplemented by the contraint arising from the equations of motion of the 
standard Weyl fields we have eliminated. This is compatible, for example, with the $\rho$ equation of motion however when we come to consider higher derivative theories the form of this contraint will 
change.\footnote{We shall discuss how we can avoid this in the remainder of this section, which is particularly useful when considering higher derivative theories.}
Alternatively one could impose the gauge fixing conditions
$ L_{\mathbf{ij}}=\pm\tfrac{L'}{\sqrt2}\delta_{\mathbf{ij}}$
where $L'$ is a non-constant scalar field, and the normalization is chosen such 
that $L^2=L_{ij}L^{ij}={L'}^2$, however in such a case the local SU(2) symmetry 
of the superconformal gravity will only have been fixed down to local U(1). 
Furthermore the necessary compensating special supersymmetry transformation to 
maintain this gauge will become dependant on $dL$. This may be an interesting 
theory, but it is somewhat different from the ungauged N-R supergravity we wish 
to construct here, and we hope to return to this in future work. 

Following \cite{Coomans:2012cf} we then find the action and 
supersymmetry 
transformations given below in (\ref{conformalsusytransDW}, 
\ref{off-shellBaction}) under the gauge fixings given in 
\eqref{conformalgaugefixing}.  We can also obtain this theory in a different 
way 
which was suggested in \cite{Fujita:2001kv}, which will be useful to generalise 
the coupling to vector multiplets and higher derivative theories in the next sections.
We introduce an additional vector multiplet 
$\mathbf{V}_{\flat}=(\rho^\flat,A^\flat_\mu,\lambda^{\flat\mathbf{i}},Y^{
\flat\mathbf{ij}})$. Combining this with a linear multiplet composed of a 
vector 
multiplet that we shall denote 
$\mathbf{V}_D=(\sigma,C_\mu,\psi^{\mathbf{i}},Y^{\mathbf{ij}})$ in the density 
formula \eqref{actionformula} we obtain a suitable Lagrangian density which we 
denote $\mathcal{L}_{\mathbf{V'}}$, and we will take the Lagrangian to be
\begin{equation}
 \mathcal{L}=\mathcal{L}_L +\mathcal{L}_{\mathbf{V'}} \ .
\end{equation}

Examining the equations of motion for the vector multiplet 
$\mathbf{V}_{\flat}=(\rho^\flat,A^\flat_\mu,\lambda^{\flat\mathbf{i}},Y^{
\flat\mathbf{ij}})$ directly in the action formula \eqref{actionformula}, since 
the composite linear multiplet does not now depend on these fields, we see that 
the fields $\mathbf{V}_\flat$ act as Lagrange multipliers, whose equations of motion set 
the fields of the composite linear multiplet to zero\footnote{Note that this 
also clearly satisfies the constraint $\mathcal{D}^mE_m=0$. } and one obtains 
expressions for the standard-Weyl multiplet matter fields in terms of 
$\mathbf{V}_D$,
 \begin{align}
 \underline{T}^{mn} &= \tfrac{\sigma^{-2}}{8}(\sigma\hat{G}^{mn} + 
\tfrac{1}{6}\epsilon^{mnpqr}\hat{H}_{pqr} + 
\tfrac{i}{4}\bar{\psi}\gamma^{mn}\psi) \ ,\nonumber\\
\underline{\chi}^{\mathbf{i}}&=\tfrac{i}{8}\sigma^{-1}\gamma^{m}\mathcal{D}
_m\psi^{\mathbf{i}} 
+\tfrac{i}{16}\sigma^{-2}\gamma^{m}(\mathcal{D}_m\sigma)\psi^{\mathbf{i}} - 
\tfrac{\sigma^{-2}}{32}\gamma_{mn}\hat{G}^{mn}\psi^{\mathbf{i}} \nonumber\\
&+ \tfrac{\sigma^{-1}}{4}\gamma_{mn}\underline{T}^{mn}\psi^{\mathbf{i}} 
+\tfrac{\sigma^{-2}}{8}\underline{Y}^{\mathbf{ij}}\psi_{\mathbf{j}} \ 
,\nonumber\\
\underline{D} &= \tfrac{1}{4\sigma}\hat{\Box}\sigma 
+\tfrac{1}{8\sigma^2}(\mathcal{D}\sigma)^2 - \tfrac{1}{16\sigma^{2}}\hat{G}^2  
+ 
\tfrac{1}{2}\underline{f}_m^m \nonumber\\
&+\left(2\sigma^{-1}\hat{G}_{mn} -\tfrac{26}{3}\underline{T}_{mn}\  
+\tfrac{i\sigma^{-2}}{4}\bar{\psi}\gamma_{mn}\psi\right)\underline{T}^{mn}
+\tfrac{1}{2}\bar{\psi}_m\gamma^m\gamma_{np}\underline{T}^{np}\psi 
\nonumber\\
&+\tfrac{1}{2}\bar{\underline{\phi}}^m\gamma_m\psi  
-\tfrac{\sigma^{-2}}{8}\bar{\psi}\gamma^m\mathcal{D}_m\psi - 
\tfrac{\sigma^{-4}}{64}\bar{\psi}^{\mathbf{i}}\psi^{\mathbf{j}}\bar{\psi}_{
\mathbf{i}}\psi_{\mathbf{j}} -4i\sigma^{-1}\psi\underline{\chi}\ , \nonumber\\
\underline{Y}^{\mathbf{ij}}&=\tfrac{i}{4}\sigma^{-1}\bar{\psi}^{\mathbf{i}}{\psi
}^{\mathbf{j}} \ ,\label{dilatonweyldefs}
\end{align}
where
\begin{eqnarray}
 \hat{G}_{\mu\nu} &=& G_{\mu\nu} - \bar{\psi}_{[\mu}\gamma_{\nu]}\psi + 
\tfrac{i}{2}\sigma\bar{\psi}_{[\mu}\psi_{\nu]} \ ,\nonumber\\
\hat{H}_{\mu\nu\rho} &=& H_{\mu\nu\rho} - 
\tfrac{3}{4}\sigma^2\bar{\psi}_{[\mu}\gamma_{\nu}\psi_{\rho]} - 
\tfrac{3i}{2}\sigma\bar{\psi}_{[\mu}\gamma_{\nu\rho]}\psi \ ,\nonumber\\
H_{\mu\nu\rho} &=& 3\partial_{[\mu} B_{\nu\rho]} 
+\tfrac{3}{2}C_{[\mu}G_{\nu\rho]} \ ,\nonumber\\
\hat{\Box}\sigma &=& (\nabla^m -2b^m)\mathcal{D}_m\sigma - 
\tfrac{i}{2}\bar{\psi}_m\mathcal{D}^m\psi 
-2\sigma\bar{\psi}_m\gamma^m\underline{\chi} \ , \label{dilatonweyldefs2}
\end{eqnarray}
and $G=dC$.
The equation of motion for $D$ now implies
$ L=\sigma^2\rho^\flat$, 
so the gauge fixing conditions \eqref{gaugefixing}
can be implemented, as the constraint arising from the $D$ equation of motion 
can be solved in terms of $\rho^\flat$ which is a Lagrange multiplier and the 
other fields of $\mathbf{V}_\flat$ can be similarly used to solve the 
$T_{mn},\chi^{\mathbf{i}}$ equations of motion. As above we use the expressions 
\eqref{dilatonweyldefs} to define a new gravitational multiplet, and will take 
them to be identities, so that the term involving the Lagrange multipliers can 
be neglected in the action, since the composite linear multiplet is now 
identically vanishing. In particular we can always solve the contraints coming from the standard weyl fields we have eliminated using the Lagrange multipliers.

 Note that in this case the contribution to the superconformal action from the 
vector multiplets is completely contained in the expressions for the previously 
independent standard-Weyl multiplet matter fields, which are now composite. If 
we take the most general contribution from the vector multiplet $\mathbf{V}_D$ 
that still allows for the $\mathbf{V}_\flat$ vector multiplet to be a Lagrange 
multiplier, i.e. we add the Lagrangian density $\mathcal{L}_{\mathbf{V}_D}$ 
formed from the three copies of $\mathbf{V}_D$ we find, using the expressions 
\eqref{dilatonweyldefs}, that 
$
 \mathcal{L}_{\mathbf{V}_D}=0. 
$
Indeed this must be the case as there are no terms in the Lagrangian density 
\eqref{actionformula} that do not involve the composite linear multiplet, 
which, 
as we have seen above, vanishes.

Let us now summarize the details of the dilaton-Weyl multiplet, which is made 
up 
of the vielbien $e_\mu^m$, gravitino $\psi_\mu^{\mathbf{i}}$, graviphoton gauge 
field $C_{\mu}$, a two-form gauge field $B_{\mu\nu}$,
the dilaton $\sigma$, the dilatino $\psi^{\mathbf{i}}$ and an auxiliary SU(2) 
triplet of vectors $V^{\mathbf{ij}}_\mu$ with 
$V^{\mathbf{ij}}_\mu=V^{\mathbf{ji}}_{\mu}$ and a gauge field for local 
dilatations $b_{\mu}$.
These transform under supersymmetry with parameter $\epsilon^{\mathbf{i}}$ and 
special supersymmetry with parameter $\eta^{\mathbf{i}}$ as
\begin{align}
 \delta e_\mu^m &=\tfrac{1}{2}\bar{\epsilon}\gamma^m\psi_\mu \ ,\nonumber\\
 \delta \psi_\mu^{\mathbf{i}} &= (\nabla_\mu 
+\tfrac{1}{2}b_\mu)\epsilon^{\mathbf{i}} 
-V_{\mu}^{\mathbf{ij}}\epsilon_{\mathbf{j}} 
+i\gamma_{mn}\underline{T}^{mn}\gamma_\mu\epsilon^{\mathbf{i}} - 
i\gamma_\mu\eta^{\mathbf{i}}\ ,\nonumber\\
 \delta V^{\mathbf{ij}}_\mu &= -\tfrac{3i}{2}\bar{\epsilon}^{\mathbf{(i}} 
\underline{\phi}^{\mathbf{j)}}_\mu +4\bar{\epsilon}^{\mathbf{(i}}\gamma_\mu 
\underline{\chi}^{\mathbf{j)}} + 
i\bar{\epsilon}^{\mathbf{(i}}\gamma_{mn}\underline{T}^{mn}\psi^{\mathbf{j)}}
_\mu 
+\tfrac{3i}{2}\bar{\eta}^{\mathbf{(i}}\psi^{\mathbf{j)}}_\mu\ , \nonumber\\
 \delta C_\mu &= -\tfrac{i}{2}\sigma\bar{\epsilon}\psi_\mu + 
\tfrac{1}{2}\bar{\epsilon}\gamma_\mu\psi \ ,\nonumber\\
 \delta B_{\mu\nu} &= 
\tfrac{1}{2}\sigma^2\bar{\epsilon}\gamma_{[\mu}\psi_{\nu]} 
+ \tfrac{i}{2}\sigma \bar{\epsilon}\gamma_{\mu\nu}\psi + 
C_{[\mu}\delta(\epsilon)C_{\nu]} \ ,\nonumber\\
 \delta \psi^{\mathbf{i}} &= 
-\tfrac{1}{4}\gamma_{mn}\hat{G}^{mn}\epsilon^{\mathbf{i}} - 
\tfrac{i}{2}\gamma^m(\mathcal{D}_m\sigma)\epsilon^{\mathbf{i}} + 
\sigma\gamma_{mn}\underline{T}^{mn}\epsilon^{\mathbf{i}} - 
\tfrac{i}{4}\sigma^{-1}\epsilon_{\mathbf{j}}\bar{\psi}^{\mathbf{i}}\psi^{\mathbf
{j}} + \sigma\eta^{\mathbf{i}}\ , \nonumber\\
 \delta \sigma &= \tfrac{i}{2}\bar{\epsilon}\psi \ ,\nonumber\\
 \delta b_\mu &= \tfrac{i}{2}\bar{\epsilon}\underline{\phi}_\mu - 
2\bar{\epsilon}\gamma_\mu\underline{\chi} + \tfrac{i}{2}\bar{\eta}\psi_\mu \ 
,\label{conformalsusytransDW}
\end{align}
where we have underlined composite fields the expressions for which are listed 
in \eqref{compositesuperconformalfields}
but now additionally $\underline{T}_{mn}$,$\underline{D}$ and 
$\underline{\chi}^{\mathbf{i}}$ are given by their expressions in 
\eqref{dilatonweyldefs}. The supercovariant field strength $\hat{H}$ defined in 
\eqref{dilatonweyldefs2} obeys the generalized Bianchi identity
\begin{equation}
\mathcal{D}_{[\mu}\hat{H}_{\nu\rho\sigma]}=\tfrac{3}{4}\hat{G}_{[\mu\nu}\hat{G}_
{\rho\sigma]} \ , \label{genbianchi1vector}
\end{equation}
 where $G=dC$.

Armed with the superconformal theory we now wish to gauge fix down to the N-R 
supergravity. First we choose 
\begin{equation}
b_\mu=0\ , \qquad L_{\mathbf{ij}}=\tfrac{L}{\sqrt{2}}\delta_{\mathbf{ij}} \ 
,\qquad \varphi^{\mathbf{i}}=0 \ .\label{conformalgaugefixing}
\end{equation}
The first condition breaks local dilatational invariance and fixes the form of 
the necessary compensating special conformal boosts, the second breaks local 
SU(2) down to U$(1)_R$, where $L$ is constant, and the third 
fixes special supersymmetry. Choosing the value of $L$ is a choice of 
dilatation. In order to maintain this gauge we must set 
\begin{equation}
 \eta_{\mathbf{k}}=\tfrac{1}{3}(\gamma\cdot T)\epsilon_{\mathbf{k}} 
-\tfrac{i}{2}(\gamma^m E_m)\delta_{\mathbf{ik}}\epsilon^{\mathbf{i}} \ 
,\label{etafixing}
\end{equation}
where in order to avoid confusion we point out that $E_m$ is the vector of the 
compensating linear multiplet, not the composite one.

Under these gauge fixing conditions we obtain for the bosonic part of the 
action\footnote{We have written the action in this way to emphasize the fact 
that the relative signs of the terms appearing are not dependent on the gauge 
fixing choice $L=\pm1$.  Rather this choice only gives an overall sign to this 
contribution to the action.}
\begin{align}
 e^{-1}L^{-1}\mathcal{L}_L&= -\tfrac{1}{2}R + \tfrac{1}{4}\sigma^{-2}G^2 
+\tfrac{1}{6}\sigma^{-4}H^2 +\tfrac{3}{2}\sigma^{-2}(d\sigma)^2 \nonumber\\
 & -{V'}^{\mathbf{ij}}_\mu {V'}_{\mathbf{ij}}^\mu - N^2 + L^{-2}P_\mu P^\mu 
+\sqrt2 L^{-2} P^\mu V_\mu \ ,\label{off-shellBaction}
\end{align}
where we have decomposed $V_\mu^{\mathbf{ij}}$ into its traceful and traceless 
parts \cite{Coomans:2012cf}
\begin{equation}
 V^{\mathbf{ij}}_\mu={V'}^{\mathbf{ij}} +\tfrac{1}{2}\delta^{\mathbf{ij}}V_\mu\ 
, \qquad {V'}^{\mathbf{ij}}\delta_{\mathbf{ij}}=0\ , \label{tracesplitV}
\end{equation}
and $P^\mu$ denotes the bosonic part of $E^\mu$. Finally we set $L=1$.
The action \eqref{off-shellBaction} is invariant under the supersymmetry 
transformations \eqref{conformalsusytransDW}, with the special supersymmetry 
parameter $\eta^{\mathbf{i}}$ replaced by its expression \eqref{etafixing}.
To arrive at the on-shell formulation we may next eliminate auxiliary fields 
$P^\mu$, $N$ and $V_\mu^{\mathbf{ij}}$ by their equations of motion which imply 
these fields vanish, and the supersymmetry transformations 
become:\footnote{Note 
that due to the equation of motion for $P^\mu$ the special supersymmetry 
parameter now reads 
$\eta^{\mathbf{i}}=\tfrac{1}{3}\gamma_{mn}T^{mn}\epsilon^{\mathbf{i}}$ if we 
ignore terms quadratic in the fermionic fields.}
\begin{align}
 \delta e_\mu^m &=\tfrac{1}{2}\bar{\epsilon}\gamma^m\psi_\mu \ ,\nonumber\\
 \delta \psi_\mu^{\mathbf{i}} &= \nabla_\mu \epsilon^{\mathbf{i}} 
+i\underline{T}^{mn}\left(\gamma_{mn}\gamma_\mu- 
\tfrac{1}{3}\gamma_\mu\gamma_{mn}\right)\epsilon^{\mathbf{i}} \ ,\nonumber\\
 \delta C_\mu &= -\tfrac{i}{2}\sigma\bar{\epsilon}\psi_\mu + 
\tfrac{1}{2}\bar{\epsilon}\gamma_\mu\psi \ ,\nonumber\\
 \delta B_{\mu\nu} &= 
\tfrac{1}{2}\sigma^2\bar{\epsilon}\gamma_{[\mu}\psi_{\nu]} 
+ \tfrac{i}{2}\sigma \bar{\epsilon}\gamma_{\mu\nu}\psi + 
C_{[\mu}\delta(\epsilon)C_{\nu]} \ ,\nonumber\\
 \delta \psi^{\mathbf{i}} &= 
-\tfrac{1}{4}\gamma_{mn}G^{mn}\epsilon^{\mathbf{i}} 
- \tfrac{i}{2}\gamma^m(\partial_m\sigma)\epsilon^{\mathbf{i}} + 
\tfrac{4}{3}\sigma\gamma_{mn}\underline{T}^{mn}\epsilon^{\mathbf{i}} \ , 
\nonumber\\
 \delta \sigma &= \tfrac{i}{2}\bar{\epsilon}\psi \ 
.\label{guagedfixedon-shellsusy}
\end{align}
 We must now perform some field and parameter redefinitions to bring the 
supersymmetry transformations to a same form as those in 
\cite{Nishino:2000cz,Nishino:2001ji}. We will take
 \begin{equation}
  \epsilon^{\mathbf{i}} = -\sqrt2{\epsilon'}^\mathbf{i}\ , \quad 
\psi^{\mathbf{i}}_\mu=-\sqrt2 {\psi'}^{\mathbf{i}}_\mu \ ,\quad 
\sigma=e^{\sigma'}\ , \quad \psi^{\mathbf{i}} = 
-\tfrac{\sqrt2}{\sqrt3}e^{\sigma'}{\chi'}^{\mathbf{i}}\ , \quad C_\mu =\sqrt2 
{A'}_\mu \ ,
 \end{equation}
  noting that the definition of the three form field strength has therefore 
changed to 
 \begin{equation}                                                                       
{G'}_{\mu\nu\rho}={H'}_{\mu\nu\rho}=3\partial_{[\mu}B_{\nu\rho]} + 
3{A'}_{[\mu}{F'}_{\nu\rho]}\ ,
\end{equation}
where ${F'}=d{A'}=\sqrt2 G$, and so from \eqref{genbianchi1vector} the Bianchi 
identity for $G$ now reads
\begin{equation}
\mathcal{\partial}_{[\mu}{G'}_{\nu\rho\sigma]}=\tfrac{3}{2}{F'}_{[\mu\nu}{F'}_{
\rho\sigma]} \ .
\end{equation}
 Dropping the primes we find the following supersymmetry transformations
\begin{align}
\delta {e_\mu}^m &= \bar{\epsilon} \gamma^m \psi_{\mu} \ , \nonumber\\
\delta \sigma &=\tfrac{i}{\sqrt{3}}\bar{\epsilon} \chi \ ,\nonumber \\
\delta {\psi_\mu}^{\mathbf{i}} &= \nabla_{\mu}\epsilon^{\mathbf{i}} + 
\tfrac{i}{6\sqrt{2}}e^{-\sigma}\Big( 
{\gamma_{\mu}}^{\rho\sigma}-4{\delta_\mu}^\rho \gamma^{\sigma}\Big) 
\epsilon^{\mathbf{i}}  {F_{\rho\sigma}}+\tfrac{1}{18}e^{-2\sigma}\Big( 
{\gamma_{\mu}}^{\rho\sigma\tau}-\tfrac{3}{2}{\delta_\mu}^\rho 
\gamma^{\sigma\tau}\Big)\epsilon^{\mathbf{i}} G_{\rho\sigma\tau} \ ,\nonumber \\
\delta {A_{\mu}} &=-\tfrac{i}{\sqrt{2}}e^{\sigma}\bar{\epsilon}\psi_{\mu } + 
\tfrac{1}{\sqrt{6}}e^{\sigma}\bar{\epsilon}\gamma_{\mu}\chi \ , \nonumber \\
\delta B_{\mu\nu} &=e^{2\sigma}\bar{\epsilon}\gamma_{[\mu}\psi_{\nu]} + 
\tfrac{i}{\sqrt{3}}e^{2\sigma}\bar{\epsilon}\gamma_{\mu\nu}\chi+ 
2{A_{[\mu\vert}}\delta {A_{\vert \nu ]}} \ ,\nonumber \\
\delta \chi^{\mathbf{i}} &= -\tfrac{1}{2\sqrt{6}} e^{-\sigma} 
\gamma^{\mu\nu}\epsilon^{\mathbf{i}} 
{F_{\mu\nu}}+\tfrac{i}{6\sqrt{3}}e^{-2\sigma}\gamma^{\mu\nu\rho}\epsilon^{
\mathbf{i}} G_{\mu\nu\rho} 
-\tfrac{\sqrt{3}i}{2}\gamma^{\mu}\epsilon^{\mathbf{i}} \partial_{\mu}\sigma\ ,
\nonumber 
\end{align}
under which the Lagrangian with bosonic part
\begin{equation*}
-e^{-1}\tfrac{1}{2}\mathcal{L}_L= \tfrac{1}{4}R  -\tfrac{3}{4}(d\sigma)^2 - 
\tfrac{1}{4}e^{-2\sigma}G^2 -\tfrac{1}{12}e^{-4\sigma}H^2 \ ,
\end{equation*}
is invariant. Taking account of the different curvature conventions between 
\cite{Nishino:2000cz,Nishino:2001ji} and 
\cite{Bergshoeff:2001hc,Coomans:2012cf,Ozkan:2013nwa} by changing the sign of 
the Ricci scalar, we have thus arrived at the pure N-R formulation 
$\mathcal{N}=2$, $d=5$ supergravity. The fermionic terms up to quadratic level 
are given in \cite{Nishino:2000cz,Nishino:2001ji} and may also be cross checked 
using the results of \cite{Coomans:2012cf}.

\section{Coupling to Abelian vector multiplets.}\label{off-shellvectorcoupled}
 The superconformal vector multiplets, labelled by capital Latin indices 
$I,J,K,\ldots$ are each 
 formed from an SU(2) triplet field $Y^{\mathbf{ij}}$, the gauge field $A_\mu$, 
a gaugino $\lambda^{\mathbf{i}}$ and a scalar $\rho$. These transform under the 
supersymmetries as
 \begin{align}
  \delta A^I_\mu &= -\tfrac{i}{2}\rho^I \bar{\epsilon}\psi_{\mu} + 
\tfrac{1}{2}\bar{\epsilon}\gamma_\mu\lambda^I \ ,\nonumber\\
  \delta Y^{I\mathbf{ij}} &= 
-\tfrac{1}{2}\bar{\epsilon}^{\mathbf{(i}}\gamma^m\mathcal{D}_m\lambda^{I\mathbf{
j)}} + 
\tfrac{i}{2}\bar{\epsilon}^{\mathbf{(i}}\gamma_{mn}{T}^{|mn|}\lambda^{\mathbf{j)
}I} - 4i\rho^I\bar{\epsilon}^{\mathbf{(i}}\chi^{\mathbf{j)}} + 
\tfrac{i}{2}\bar{\eta}^{\mathbf{(i}}\lambda^{\mathbf{j)}I} \ ,\nonumber\\
  \delta \lambda^{\mathbf{i}I} &= -\tfrac{1}{4}\gamma_{mn}\hat{F}^{Imn} 
{\epsilon}^{\mathbf{i}} - 
\tfrac{i}{2}\gamma^m\mathcal{D}_m\rho^I\epsilon^{\mathbf{i}} +\rho^I 
\gamma_{mn}{T}^{mn}\epsilon^{\mathbf{i}} - 
Y^{I\mathbf{ij}}\epsilon_{\mathbf{j}} 
+\rho^I\eta^{\mathbf{i}}\ , \nonumber\\
  \delta \rho^I &=\tfrac{i}{2}\bar{\epsilon}\lambda^I \ 
,\label{off-shellvectorsusy}
 \end{align}
 where
 \begin{align}
  \mathcal{D}_\mu \rho^I &= (\partial_\mu -b_\mu)\rho^I 
-\tfrac{i}{2}\bar{\psi}_\mu\lambda^I \ ,\nonumber\\
  \mathcal{D}_\mu \lambda^{I\mathbf{i}} &= (\nabla_\mu -\tfrac{3}{2}b_\mu) 
\lambda^{I\mathbf{i}} -V_\mu^{\mathbf{ij}}\lambda^I_{\mathbf{j}} + 
\tfrac{1}{4}\gamma_{mn}\hat{F}^{Imn}\psi^{\mathbf{i}}_\mu + 
\tfrac{i}{2}\gamma^m(\mathcal{D}_m\rho^I)\psi^{\mathbf{i}}_\mu \nonumber\\
  & +Y^{I\mathbf{ij}}\psi_{\mu\mathbf{j}} - 
\rho^I\gamma_{mn}\underline{T}^{mn}\psi^{\mathbf{i}}_\mu 
-\rho^I\underline{\phi}^{\mathbf{i}}_\mu\ ,\nonumber\\
  \hat{F}^I_{\mu\nu}&=F^I_{\mu\nu} -\bar{\psi}_{[\mu}\gamma_{\nu]}\lambda^I 
+\tfrac{i}{2}\rho^I\bar{\psi}_{[\mu}\psi_{\nu]}\ ,
 \end{align}
 and where $F^I=dA^I$.
 
 These are embedded into a linear multiplet
 \begin{align}
 L(\mathbf{V})_{\mathbf{ij}} &= a_{IJ} \left( 2\rho^I Y^J_{\mathbf{ij}} 
-\tfrac{i}{2}\bar{\lambda}^I_{\mathbf{i}}\lambda^J_{\mathbf{j}}\right) \ 
,\nonumber\\
 \varphi_{\mathbf{i}}(\mathbf{V}) &= 
a_{IJ}\left(i\rho^I\gamma^m\mathcal{D}_m\lambda^J_{\mathbf{i}} + 
2\rho^I\gamma_{mn}T^{mn}\lambda^J_{\mathbf{i}} 
-8\rho^{I}\rho^{J}\chi_{\mathbf{i}} \right.\nonumber\\
 &\left.- \tfrac{1}{4}\gamma^{mn}\hat{F}^I_{mn}\lambda^J_{\mathbf{i}} + 
\tfrac{i}{2}\gamma^m(\mathcal{D}_m\rho^I)\lambda^J_{\mathbf{i}} 
-Y^I_{\mathbf{ij}}\lambda^{J\mathbf{j}} \right)\ ,\nonumber\\
 E^m(\mathbf{V}) &= a_{IJ}\left( \mathcal{D}_n\left( -\rho^I \hat{F}^{Jmn} 
+8\rho^I\rho^J T^{mn} -\tfrac{i}{4}\bar{\lambda}^{I}\gamma^{mn}\lambda^{J} 
\right) -\tfrac{1}{8}\epsilon^{mnpqr}F^{I}_{np}F^{J}_{qr} \right)\ , \nonumber\\
 N(\mathbf{V}) &= a_{IJ} \left(\rho^I \Box \rho^J + \tfrac{1}{2}(\mathcal{D}_m 
\rho^I) (\mathcal{D}^m\rho^J) - \tfrac{1}{4}\hat{F}^I_{mn}\hat{F}^{Jmn} 
+Y^{I\mathbf{ij}}Y^J_{\mathbf{ij}} +8\rho^I\hat{F}^J_{mn}T^{mn} \right. 
\nonumber\\
 &\left. - 4\rho^I\rho^J\left(D+\tfrac{26}{3}T^2\right) 
-\tfrac{1}{2}\bar{\lambda^I}\gamma^m\mathcal{D}_m\lambda^J 
+i\bar{\lambda}^I\gamma_{mn}T^{mn}\lambda^J + 16i\rho^I\bar{\chi}\lambda^J 
\right)\ ,  \label{compositelinearmultiplet}
\end{align}
where $a_{IJ}$ is a symmetric constant matrix.
 We then compose a density from these with the Lagrange multiplier vector 
multiplet, $\mathbf{V}_\flat$, and solve the equations of motion of the fields 
of $\mathbf{V}_\flat$ in terms of the standard-Weyl fields, i.e we get the 
equations
 $L(\mathbf{V})=0, E^a(\mathbf{V})=0, \varphi^{\mathbf{i}}(\mathbf{V})=0$ and 
$N(\mathbf{V})=0$. Similarly to the above we can implement the $D$ equation as 
a 
constraint by defining a new extended dilaton-Weyl multiplet containing the 
vector fields and solving the constraints for the Lagrange multipliers. 
 
 Note we can diagonalize $a_{IJ}$ using a constant GL$(n,\mathbb{R})$ 
transformation, which is just a constant linear field redefinition of the 
vector 
multiplets. Furthermore we can set the diagonal entries to be 
$\pm1$.\footnote{We shall assume for the time being that $\det{a}\neq0$, 
however 
it is clear that as we may still diagonalize $a_{IJ}$ the vector multiplet 
directions for which $a_{IJ}$ has zero eigenvalues will not contribute to this 
action.} We shall take a Lorentzian signature, 
$\eta_{IJ}=\mathrm{diag}(-1,1,\ldots,1)$, so that we arrive at the N-R 
formulation, which is presumably needed to ensure the absence of 
ghosts.\footnote{Note that we have not analysed fully whether there is any way 
to avoid the introduction of ghosts in different signatures for the matrix 
$a_{IJ}$, but we expect that the Lorentzian signature is necessary.} 
 
 Defining $\mathcal{A}
=\eta_{IJ}\rho^I\rho^J$, $\mathcal{A}_I=\eta_{IJ}\rho^J$  and solving the 
equations of motion for the Lagrange multiplier vector multiplet we obtain

  \begin{align}
\mathcal{A}\underline{T}^{mn} &=   
-\tfrac{1}{8}\left(\tfrac{1}{6}\epsilon^{mnpqr}\hat{H}_{pqr}  
-\mathcal{A}_I\hat{F}^{Imn}- 
\eta_{IJ}\tfrac{i}{4}\bar{\lambda}^I\gamma^{mn}\lambda^J\right)\ ,\nonumber\\
\mathcal{A}\underline{\chi}^{\mathbf{i}} &= \eta_{IJ}\left( 
\tfrac{i}{8}\rho^I\gamma^{m}\mathcal{D}_m\lambda^{J\mathbf{i}} 
+\tfrac{i}{16}\gamma^{m}(\mathcal{D}_m\rho^I)\lambda^{J\mathbf{i}} 
\right.\nonumber\\
&\left.- \tfrac{1}{32}\gamma_{mn}\hat{F}^{Imn}\lambda^{J\mathbf{i}} + 
\tfrac{1}{4}\rho^I\gamma_{mn}\underline{T}^{mn}\lambda^{J\mathbf{i}} 
-\tfrac{1}{8}\underline{Y}^I_{\mathbf{ij}}\lambda^{J\mathbf{j}}\right)\ , 
\nonumber\\
\mathcal{A}\underline{D} &= -\tfrac{26}{3}\mathcal{A}\underline{T}^2 + 
\eta_{IJ} 
\left( \tfrac{1}{4}\rho^{I}\Box\rho^J 
+\tfrac{1}{8}(\mathcal{D}\rho^I)(\mathcal{D}\rho^J) - 
\tfrac{1}{16}\hat{F}_{mn}^I\hat{F}^{Jmn} 
-\tfrac{1}{8}\bar{\lambda}^I\gamma^m\mathcal{D}_m\lambda^J \right.\nonumber\\
&\left.+ \tfrac{1}{4}\underline{Y}^I_{\mathbf{ij}}\underline{Y}^{J\mathbf{ij}} 
-4i\rho^{I}\lambda^{J}\underline{\chi}  +\left(2\rho^I\hat{F}^J_{mn}  
+\tfrac{i}{4}\bar{\lambda}^I\gamma_{mn}\lambda^J \right)\underline{T}^{mn} 
\right)  \ ,\nonumber\\ 
\mathcal{A}_{I}\underline{Y}^{\mathbf{ij}I}&=\tfrac{i}{4}\eta_{IJ}\bar{\lambda}^
{I\mathbf{i}}\lambda^{J\mathbf{j}} \ ,\label{vanishinglmconditions}
 \end{align}
 where
 \begin{eqnarray}
  \Box\rho^I &=& (\nabla^m -2b^m)\mathcal{D}_m\rho^I - 
\tfrac{i}{2}\bar{\psi}_m\mathcal{D}^m\lambda^ 
I-2\rho^I\bar{\psi}_m\gamma^m\underline{\chi} 
  +\tfrac{1}{2}\bar{\psi}_m\gamma^m\gamma_{np}\underline{T}^{np}\lambda^I 
\nonumber\\
  &&+\tfrac{1}{2}\bar{\underline{\phi}}^m\gamma_m\lambda^I + 
2{\underline{f}_m}^m\rho^I  \ .
 \end{eqnarray}

We will interpret the last equation in \eqref{vanishinglmconditions} as a 
definition for $\underline{Y}^0_{\mathbf{ij}}$ in terms of the fields of the 
dilaton-Weyl multiplet and the remaining vector multiplets, which is why we 
have 
underlined $Y^{I}_{\mathbf{ij}}$ in the above expressions. We have introduced 
the three form $H$
 \begin{eqnarray}
H_{\mu\nu\rho} &=& 3\partial_{[\mu} B_{\nu\rho]} 
-\tfrac{3}{2}\eta_{IJ}A^I_{[\mu}F^J_{\nu\rho]}\ , \nonumber\\
\hat{H}_{\mu\nu\rho}&=&H_{\mu\nu\rho} 
+\tfrac{3}{4}\mathcal{A}\bar{\psi}_{[\mu}\gamma_\nu\psi_{\rho]} + 
\tfrac{3i}{2}\mathcal{A}_I\bar{\psi}_{[\mu}\gamma_{\nu\rho]}\lambda^I \ ,
 \end{eqnarray}
 with modified Bianchi identity
\begin{equation} 
\nabla_{[m}\hat{H}_{npq]}=-\tfrac{3}{4}\eta_{IJ}\hat{F}^I_{[mn}\hat{F}^J_{pq]}\ 
,
\end{equation}
in order to solve the composite linear multiplet vector equation, $E^m=0$.
The two form gauge field $B_{\mu\nu}$ transforms under supersymmetry as 
\begin{equation}
 \delta 
B_{\mu\nu}=-\tfrac{1}{2}\mathcal{A}\bar{\epsilon}\gamma_{[\mu}\psi_{\nu]} - 
\tfrac{i}{2}\mathcal{A}_I\bar{\epsilon}\gamma_{\mu\nu}\lambda^I - 
\eta_{IJ}A^I_{[\mu}\delta A^J_{\nu]}\ ,
\end{equation}
and the gauge invariance of $H$ implies a suitable gauge transformation of $B$ 
is
\begin{equation}
\delta B_{\mu\nu} = 
2\partial_{[\mu}\Lambda_{\nu]}+\tfrac{1}{2}\eta_{IJ}\Lambda^I F^J_{\mu\nu} \ .
\end{equation}
We summarize the general dilaton-Weyl multiplets we have constructed following 
\cite{Fujita:2001kv}, in which this enlarged algebra was shown to close 
off-shell, in the conventions of \cite{Coomans:2012cf,Ozkan:2013nwa} in 
appendix 
\ref{gendilatonweyl}.

Inserting these expressions into \eqref{LAGLstandard} and performing the gauge 
fixing \eqref{gaugefixing}, and setting $L=1$ we obtain for the bosonic part of 
the off-shell Lagrangian\footnote{Note that the only terms that will change 
with 
respect to the Lagrangian of the pure case are those involving D and T.}
\begin{eqnarray}
 e^{-1}\mathcal{L}_L&=& -\tfrac{1}{2}R -{V'}^{\mathbf{ij}}_\mu 
{V'}_{\mathbf{ij}}^\mu - N^2 + P_\mu P^\mu +\sqrt2 P^\mu V_\mu \nonumber\\
 &&- \tfrac{1}{4}\mathcal{A}^{-1}a_{IJ}F^I\cdot F^J 
+\tfrac{1}{2}\mathcal{A}^{-2}\mathcal{A}_I\mathcal{A}_JF^I\cdot F^J 
+\mathcal{A}^{-1}\eta_{IJ}\underline{Y}^{I\mathbf{ij}}\underline{Y}_{\mathbf{ij}
}^J \nonumber\\ 
 &&-\tfrac{1}{2}\mathcal{A}^{-1}a_{IJ}(d\rho^I)\cdot (d\rho^J) 
+\mathcal{A}^{-2}\mathcal{A}_I\mathcal{A}_J(d\rho^I)\cdot(d\rho^J)
 -\tfrac{1}{6}\mathcal{A}^{-2}H^2 \ ,
 \label{off-shellvectorcoupledLaction}
\end{eqnarray}
where we have yet to implement the identity involving $Y^{I\mathbf{ij}}$ coming 
from the last equation in \eqref{vanishinglmconditions} in this Lagrangian.

Next we make a non-constant redefinition of the scalar fields
\begin{equation}
\sigma' = \tfrac{1}{2}\ln (-\mathcal{A}) \ ,\qquad {\rho'}^i = 
(-\mathcal{A})^{-\tfrac{1}{2}}\rho^i \ , \label{scalartrans}
\end{equation}
so inverting this we get
\begin{equation}
 \mathcal{A}=-e^{2\sigma'}\ , \qquad {\rho}^i = e^{\sigma'}{\rho'}^i \qquad 
\implies \qquad \rho^0=e^{\sigma'}\sqrt{1 +\delta_{ij}{\rho'}^i{\rho'}^j} := 
e^{\sigma'}L^0 \label{L0def} \ .
\end{equation}
Note that as $\mathcal{A}$ is just a quadratic polynomial of the fields 
$\rho^I$ 
it is continuous. We shall assume it never vanishes, otherwise our definitions 
for the standard-Weyl multiplet fields we have eliminated become singular. Thus 
we shall take the case $\mathcal{A}<0$ in what follows, and the positive case 
could be treated identically changing the sign of $\mathcal{A}$ in the 
transformation, although this appears to change the signature of the scalar 
manifold and would therefore introduce ghosts. Note that this transformation is 
a well defined coordinate transformation for the subspace $\mathcal{A}<0$.
Similarly we transform the gauginos such that
\begin{equation}
 \lambda^{0\mathbf{i}}=e^{\sigma'}L^0{\lambda'}^{0\mathbf{i}} 
+e^{\sigma'}(L^0)^{-1}{\rho'}^i\delta_{ij}{\lambda'}^{j\mathbf{i}} \ ,\qquad 
\lambda^{i\mathbf{i}}=e^{\sigma'}({\lambda'}^{i\mathbf{i}} 
+{\rho'}^i{\lambda'}^{0\mathbf{i}})\ , \label{gauginotrans}
\end{equation}and the inverse of this transformation is
\begin{eqnarray}
 {\lambda'}^{0\mathbf{i}}&=&-\mathcal{A}^{-1}\mathcal{A}_I\lambda^{I\mathbf{i}} 
\ ,\nonumber\\
{\lambda'}^{i\mathbf{i}}&=&\frac{1}{\sqrt{-\mathcal{A}}}\left(\lambda^{i\mathbf{
i}}+\rho^i\mathcal{A}^{-1}\mathcal{A}_I\lambda^{I\mathbf{i}}\right)\ . 
\label{gauginotransinverse}
\end{eqnarray}
We leave all other fields fixed. Note that after this transformation the 
condition on the auxiliary fields $Y^{I\mathbf{ij}}$ from 
\eqref{vanishinglmconditions} becomes
\begin{equation}
Y^{0\mathbf{ij}}=(L^0)^{-1}\delta_{ij}{\rho'}^{i}Y^{j\mathbf{ij}} \ . 
\label{Y0condition}
\end{equation}
So dropping the primes, the bosonic part of the resulting off-shell Lagrangian 
is
\begin{eqnarray}
 e^{-1}\mathcal{L}_L&=& -\tfrac{1}{2}R -{V'}^{\mathbf{ij}}_\mu 
{V'}_{\mathbf{ij}}^\mu - N^2 + P_\mu P^\mu +\sqrt2 P^\mu V_\mu \nonumber\\
 &&+ \tfrac{1}{4}e^{-2\sigma}\left(\eta_{IJ}+2L_IL_J\right)F^I\cdot F^J 
 +e^{-2\sigma}\eta_{IJ}L^I_\alpha 
L^J_\beta\delta^\alpha_i\delta^\beta_j{Y}^{i\mathbf{ij}}{Y}_{\mathbf{ij}}^j 
\nonumber\\ 
 &&+\tfrac{3}{2}(d\sigma)\cdot(d\sigma) +\tfrac{1}{2}\eta_{IJ}L^I_\alpha 
L^J_\beta\delta^\alpha_i\delta^\beta_j(d\rho^i)\cdot(d\rho^j)
 +\tfrac{1}{6}e^{-4\sigma}H^2 \ ,
 \label{off-shelltransformedaction}
\end{eqnarray}
where we have defined
\begin{equation}
 L^I=(L^0,\rho^i)\ , \qquad  L^0_\alpha=(L^0)^{-1}\delta_{\alpha i}\rho^i \ 
,\qquad L^i_\alpha=\delta^i_\alpha\ ,
 \end{equation}
and 
 \begin{equation}
 L_I=(L^0,-\delta_{ij}\rho^j) \ ,\qquad L^\alpha_0= 
-(L^0)\delta_i^{\alpha}\rho^i   \ ,\qquad L_i^\alpha=\delta_i^\alpha+  
\delta_{ij}\delta^{\alpha}_k\rho^j\rho^k\ ,
\end{equation}
and where $L^0$ is defined by \eqref{L0def}.
One may check that after introducing indices $A=(0,a)$ which are raised and 
lowered with the metric $\eta_{AB}=\textrm{diag}(-,+,\cdots,+)$ and 
identifying\footnote{Note that with this definition 
$\eta_{IJ}L^IL^J=-L_IL^I=-1$.} $L_0^I=L^I$ and $L^0_I=L_I$ we have
\begin{equation}
L^I_AL^A_J:=L^IL_J+L^I_aL_J^a=\delta_J^I\ , \qquad L_I^A L^I_B=\delta^A_B \ 
,\label{Lsorthogonal}
\end{equation}
independently of the frame, as long as the vielbein are invertible. Of course 
we 
also need these vielbein in order to define the fermions locally on the scalar 
manifold.
Explicitly the vielbein $V^a_\alpha$ is
\begin{equation}
 V^a_\alpha=\delta_\alpha^a-\frac{1}{L^0(L^0+1)}\delta^{a}_{i}\delta_{\alpha 
j}\rho^{i}\rho^j \  ,
\end{equation}
with inverse
\begin{equation}
 V^\alpha_a=\delta^\alpha_a + \frac{1}{(L^0+1)}\delta^{\alpha}_{i}\delta_{a 
j}\rho^{i}\rho^j\  .
\end{equation}
We note that this means that the transformations 
(\ref{gauginotrans}),(\ref{gauginotransinverse}) may be written
\begin{equation}
 {\lambda}^{I\mathbf{i}}=e^{\sigma'}(L^I{\lambda'}^{0\mathbf{i}} + 
L^I_\alpha\delta^\alpha_i{\lambda'}^{i\mathbf{i}})= 
e^{\sigma'}L^I_A{\lambda'}^{A\mathbf{i}}\ ,\qquad
 {\lambda'}^{A\mathbf{i}}=\tfrac{1}{\sqrt{-\mathcal{A}}}L^A_I \lambda^I\ ,
\end{equation}
where we defined ${\lambda'}^A=\delta^A_J{\lambda'}^J$ and that the condition 
\eqref{Y0condition} implies
\begin{equation}
 Y^{I\mathbf{ij}}=L^I_aV^a_\alpha \delta^\alpha_iY^{i\mathbf{ij}}\ .
\end{equation}

To maintain our gauge fixing recall the special supersymmetry parameter must be 
set to
\begin{equation}
\eta_{\mathbf{k}}=\tfrac{1}{3}(\gamma\cdot T)\epsilon_{\mathbf{k}} 
-\tfrac{i}{2}(\gamma^m P_m)\delta_{\mathbf{ik}}\epsilon^{\mathbf{i}} + \ldots\ 
, 
\label{etafixedquad}
\end{equation}
where $\ldots$ signifies terms higher order in the fermions.
The supersymmetry transformations become
\begin{align}
 \delta e_\mu^m &=\tfrac{1}{2}\bar{\epsilon}\gamma^m\psi_\mu \ ,\nonumber\\
 \delta \psi_\mu^{\mathbf{i}} &= \nabla_\mu \epsilon^{\mathbf{i}} 
-V_{\mu}^{\mathbf{ij}}\epsilon_{\mathbf{j}} 
+i(\gamma_{mn}\gamma_\mu-\tfrac{1}{3}\gamma_\mu\gamma_{mn}){\underline{T}}^{mn}
\epsilon^{\mathbf{i}} - 
\tfrac{1}{2}\gamma_\mu\gamma_mP^m\epsilon^{\mathbf{ij}}\delta_{\mathbf{jk}}{
\epsilon}^{\mathbf{k}}\ ,\nonumber\\
 \delta V^{\mathbf{ij}}_\mu &= -\tfrac{3i}{2}\bar{\epsilon}^{\mathbf{(i}} 
\underline{\phi}^{\mathbf{j)}}_\mu +4\bar{\epsilon}^{\mathbf{(i}}\gamma_\mu 
\underline{\chi}^{\mathbf{j)}} + 
i\bar{\epsilon}^{\mathbf{(i}}\gamma_{mn}{\underline{T}}^{mn}\psi^{\mathbf{j)}}
_\mu +\tfrac{3i}{2}\bar{\underline{\eta}}^{\mathbf{(i}}\psi^{\mathbf{j)}}_\mu \ 
,\nonumber\\
 \delta A^I_\mu &= -\tfrac{i}{2}e^{\sigma}L^I \bar{\epsilon}\psi_{\mu} + 
\tfrac{1}{2}e^{\sigma}\bar{\epsilon}\gamma_\mu L^I \lambda^0 + 
\tfrac{1}{2}e^{\sigma}\bar{\epsilon}\gamma_\mu L^I_a\lambda^a\ ,\nonumber\\
 \delta B_{\mu\nu} &= 
\tfrac{1}{2}e^{2\sigma}\bar{\epsilon}\gamma_{[\mu}\psi_{\nu]} + 
\tfrac{i}{2}e^{2\sigma} \bar{\epsilon}\gamma_{\mu\nu}\chi - 
\eta_{IJ}A^I_{[\mu}\delta(\epsilon)A^J_{\nu]} \ ,\nonumber\\
 \delta Y^{i\mathbf{ij}} &= 
-\tfrac{1}{2}\bar{\epsilon}^{\mathbf{(i}}\gamma^m\mathcal{D}_m(e^{\sigma}
L^i_A\lambda^{\mathbf{j)}A}) + 
\tfrac{i}{2}e^{\sigma}L^i_A\bar{\epsilon}^{\mathbf{(i}}\gamma_{mn}\underline{T}^
{|mn|}\lambda^{\mathbf{j)}A} - 
4i\rho^i\bar{\epsilon}^{\mathbf{(i}}\underline{\chi}^{\mathbf{j)}} + 
\tfrac{i}{2}e^{\sigma}L^i_A\bar{\underline{\eta}}^{\mathbf{(i}}\lambda^{\mathbf{
j)}A} \ ,\nonumber\\
  \delta \lambda^{\mathbf{i}0} &= -\tfrac{1}{4}e^{-\sigma}L_I{F}^{Imn} 
{\gamma_{mn}\epsilon}^{\mathbf{i}} - 
\tfrac{i}{2}\gamma^m(\partial_m\sigma)\epsilon^{\mathbf{i}} 
  + \tfrac{4}{3}\gamma_{mn}\underline{T}^{mn}\epsilon^{\mathbf{i}}  \ 
,\nonumber\\
  \delta \lambda^{\mathbf{i}a} &= 
-\tfrac{1}{4}e^{-\sigma}L_I^a{F}^{Imn}\gamma_{mn} {\epsilon}^{\mathbf{i}} 
  - \tfrac{i}{2}\gamma^m 
V^a_\alpha(\partial_m\varphi^\alpha)\epsilon^{\mathbf{i}} 
  - e^{-\sigma}V^a_\alpha\delta^\alpha_iY^{i\mathbf{ij}}\epsilon_{\mathbf{j}}  
\ 
,\nonumber\\
  \delta \sigma &=\tfrac{i}{2}\bar{\epsilon}\lambda^0\ ,\nonumber\\
  \delta \rho^i &=\tfrac{i}{2}\bar{\epsilon}\delta_\alpha^i 
V^{\alpha}_a\lambda^a \ ,\label{off-shellsusytranspoincare}
 \end{align}
where we have underlined composite fields, explicit expressions for which are 
given in (\ref{compositesuperconformalfields2}, 
\ref{compositesuperconformalfields3}, \ref{etafixedquad}).

The action \eqref{off-shelltransformedaction} with supersymmetry 
transformations 
\eqref{off-shellsusytranspoincare} is an off-shell version of the N-R 
supergravity presented in \cite{Nishino:2000cz,Nishino:2001ji} which was first 
described in \cite{Fujita:2001kv}\footnote{We add to that work here by 
describing fully the scalar manifold, identifying the dilaton from regularity 
of 
the supersymmetry transformations and giving an explicit map to the conventions 
of \cite{Nishino:2000cz,Nishino:2001ji}.}.

Next we relabel the scalars $\varphi^\alpha=\rho^i\delta^\alpha_i$, and 
introducing the associated (n-1)-dimensional Riemannian metric 
$g_{\alpha\beta}$ 
we find
\begin{equation}
 g_{\alpha\beta}=\delta_{\alpha\beta} 
-\tfrac{1}{(L^0)^2}\delta_{\alpha\delta}\delta_{\beta\gamma}
\varphi^\delta\varphi^{\gamma}\ ,
\end{equation}
with inverse
\begin{equation}
  g^{\alpha\beta}=\delta^{\alpha\beta} +\varphi^\alpha\varphi^{\beta}\ , 
\label{inversescalarmetric}
\end{equation}
with $L^0=\sqrt{1+\delta_{\alpha\beta}\varphi^\alpha\varphi^\beta}$.
Next considering the tensor
\begin{equation}
 L_{IJ}=-L_IL_J+L_I^\alpha L_J^\beta 
g_{\alpha\beta}=\eta_{AB}L^A_IL^B_J=-L_IL_I+L_I^aL_J^b\delta_{ab} \,
\end{equation}
we see that\begin{equation}
L_{IJ}=\eta_{IJ}=\textrm{diag}(-,+,\cdots,+).
\end{equation}
We find that in our frame
\begin{eqnarray}
 L^I=(L^0,\delta^i_\alpha\varphi^\alpha) \ ,&  L^0_a=\delta_{a 
\alpha}\varphi^\alpha \ ,& L^i_a=\delta^i_a + 
\tfrac{1}{(L^0+1)}\delta_{a\alpha}\delta_\beta^i\varphi^\alpha\varphi^\beta\ 
,\nonumber\\
 L_I=(L^0,-\delta_{i\alpha}\varphi^\alpha) \ ,& L^a_0= 
-\delta^a_\alpha\varphi^\alpha    \ ,& 
L_i^a=\delta_i^a+\tfrac{1}{(L^0+1)}\delta^a_\alpha\delta_{\beta}
^i\varphi^\alpha\varphi^\beta\ .
\end{eqnarray}
Now using the Cartan equation
\begin{equation}
 dV^a + {A^a}_bV^b=0\ ,
\end{equation}
we may read off the connection 1-form ${A^a}_b$
and verify the differential identities
\begin{equation}
{L_A}^I\partial_{\alpha}{L_{I}}^B=\frac{1}{2}{A_\alpha}^{ab}{( 
H_{ab})_A}^{B}+{V_\alpha}^a {(K_a)_A}^B \ ,\label{cosetdiffidentities}
\end{equation}
where
\begin{equation}
{(H_{ab})_c}^d=\delta_{ac}{\delta_b}^d-\delta_{bc}{\delta_a}^d \qquad 
{(K_a)}_{0b}=\frac{1}{\xi}\delta_{ab}\ ,
\end{equation}
and with our current conventions we find $\xi=-1$. Following 
\cite{Nishino:2000cz,Nishino:2001ji}
we then find
\begin{equation}
D_{\alpha}L_I = \partial_{\alpha} L_I = \frac{1}{\xi}{L_I}^a V_{\alpha a}\ , 
\qquad D_{\alpha}{L_I}^a=\frac{1}{\xi}L_I {V_\alpha}^a\ .
\end{equation}
 Moreover, given that
 \begin{equation}
 L_{IJ}L^J = - L_I \ ,\qquad L_{IJ}{L_a}^{J}=L_{Ia}\ ,
 \end{equation}
we can evaluate the commutator
\begin{equation}
[D_\alpha,D_{\beta}]{L_I}^a = -\frac{1}{\xi^2}( 
{V_\alpha}^a{V_\beta}^b-{V_\beta}^a{V_\alpha}^b)L_{Ib}\ ,
\end{equation}
so we can read off the curvature tensor of $H^n$ which is
\begin{equation}
{R_{\alpha\beta}}^{ab}=-\frac{1}{\xi^2}( 
{V_\alpha}^a{V_\beta}^b-{V_\beta}^a{V_\alpha}^b)\ ,
\end{equation}
and the Ricci scalar is negative and given by $R=-n(n-1)/\xi^2$. 

In particular we have
 the coset algebra for the coset generators $K_a$ and $SO(n)$ generators 
$H_{ab}$, 
\begin{align}
\Big[ H_{ab}, H_{cd} 
\Big]&=\delta_{bc}H_{ad}-\delta_{ac}H_{bd}+\delta_{ad}H_{bc}-\delta_{bd}H_{ac} 
\ 
,\nonumber \\
\Big[ H_{ab}, K_c \Big]&=\delta_{bc}K_{a}-\delta_{ac}K_{b} \ ,\qquad \qquad 
[K_a,K_b]=\frac{1}{\xi^2}H_{ab}\ .
\label{genscoset}
\end{align}
So the scalar manifold is simply the coset space SO(1,n)/SO(n).

Note that the conditions \eqref{Lsorthogonal} are identities and not 
constraints, which is somewhat different to the case of vector multiplets in 
the 
background of the standard-Weyl multiplet, where the scalar field $D$ acts as a 
Lagrange multiplier to implement the very special geometry constraint in two 
derivative theories.\footnote{One can see that we do have one field that can 
act 
in this way, which is the scalar field of the compensating linear multiplet. 
It is possible that the scalar manifolds for physical tensor multiplet scalars 
in the background of the dilaton-Weyl gravitational multiplet will be similar 
to 
those of the vector multiplets in the background of the standard-Weyl 
multiplet.} Here though the constraint coming from the $D$ equation of motion 
is 
avoided by moving to the dilaton-Weyl multiplet and solving for the Lagrange 
multipliers which no longer occur in the action. 

Integrating out $V^{\mathbf{ij}}_\mu, V_\mu, P_\mu, N$ and $Y^{i\mathbf{ij}}$ 
the action for the linear multiplet becomes
\begin{eqnarray}
-e^{-1}\mathcal{L}_L&=& +\tfrac{1}{2}R - 
\tfrac{1}{4}e^{-2\sigma}(L_IL_J+L_I^aL_{Ja})F^I\cdot F^J  
 -\tfrac{1}{6}e^{-4\sigma}H^2 \nonumber\\
 &&
 -\tfrac{3}{2}(d\sigma)^2 - 
\tfrac{1}{2}g_{\alpha\beta}(d\varphi^\alpha)\cdot(d\varphi^\beta) \ 
,\label{on-shellbetterconventions}
 \end{eqnarray}
 and the supersymmetry variations are now
\begin{eqnarray}
 \delta {e}_\mu^m &=&\tfrac{1}{2}\bar{\epsilon}\gamma^m\psi_\mu \ ,\nonumber\\
 \delta \psi_\mu^{\mathbf{i}} &=& \nabla_\mu\epsilon^{\mathbf{i}} 
+i\gamma_{\nu_1\nu_2}\underline{T}^{\nu_1\nu_2}\gamma_\mu\epsilon^{\mathbf{i}} 
- 
\tfrac{i}{3}\gamma_\mu\gamma_{\nu_1\nu_2}\underline{T}^{\nu_1\nu_2}\epsilon^{
\mathbf{i}}\ ,\nonumber\\
   \delta A^I_\mu &=& -\tfrac{i}{2}e^{\sigma}L^I\bar{\epsilon}\psi_\mu + 
\tfrac{1}{2}e^{\sigma}L^I\bar{\epsilon}\gamma_\mu\chi + 
\tfrac{1}{2}e^{\sigma}\bar{\epsilon}\gamma_\mu L^I_a\lambda^a\ ,\nonumber\\
 \delta B_{\mu\nu} &=& 
\tfrac{1}{2}e^{2\sigma}\bar{\epsilon}\gamma_{[\mu}\psi_{\nu]} + 
\tfrac{i}{2}e^{2\sigma} \bar{\epsilon}\gamma_{\mu\nu}\chi - 
\eta_{IJ}A^I_{[\mu}\delta(\epsilon)A^J_{\nu]}\ , \nonumber\\
 \delta \chi^{\mathbf{i}} &=& 
-\tfrac{1}{12}\gamma_{mn}e^{-\sigma}L_I{F}^{Imn}\epsilon^{\mathbf{i}} - 
\tfrac{i}{2}\gamma^\mu(d\sigma)_\mu\epsilon^{\mathbf{i}} + 
\tfrac{i}{18}e^{-2\sigma}H^{mnp}\gamma_{mnp}\epsilon^{\mathbf{i}} \ ,
  \nonumber\\
   \delta \lambda^{\mathbf{i}a} &=& 
-\tfrac{1}{4}e^{-\sigma}L_I^a{F}^{Imn}\gamma_{mn} {\epsilon}^{\mathbf{i}} 
  - \tfrac{i}{2}\gamma^m 
V^a_\alpha(\partial_m\varphi^\alpha)\epsilon^{\mathbf{i}}  \ ,\nonumber\\
 \delta \sigma &=& \tfrac{i}{2}\bar{\epsilon}\chi  \ ,\nonumber\\
 \delta \varphi^\alpha &=& \tfrac{i}{2}\bar{\epsilon}V^\alpha_a\lambda^a \ 
.\label{susyon-shellbetterconventions}
 \end{eqnarray}

 Note that there are still some differences between this formulation and the 
N-R 
supergravity presented in \cite{Nishino:2000cz,Nishino:2001ji}, in particular 
here the parameter $\xi=-1$, whereas in \cite{Nishino:2000cz,Nishino:2001ji} 
$\xi=-\tfrac{1}{\sqrt2}$. However the differences are merely due to 
conventions, 
and the explicit (constant) field redefinition is given in appendix 
\ref{explictredef}. We find it useful to keep these conventions, as we will be 
interested in adding high derivative terms which are simple generalizations of 
those presented in \cite{Ozkan:2013nwa}.

Note that the on-shell theory with action \eqref{on-shellbetterconventions} is 
invariant under the scaling symmetry
\begin{equation} 
\sigma \rightarrow \sigma+c \qquad B_{\mu\nu}\rightarrow e^{2c}B_{\mu\nu} \qquad 
A^I_{\mu} 
\rightarrow e^{c}A^I_{\mu} \qquad G_{\mu\nu\rho}\rightarrow e^{2c} 
G_{\mu\nu\rho}\ ,\label{shiftsymmetry}
\end{equation}
and the off-shell theory \eqref{off-shelltransformedaction} with supersymmetry 
transformations \eqref{off-shellsusytranspoincare} maintains this symmetry if 
we 
also scale
\begin{equation}
 Y^{I\mathbf{ij}} \rightarrow e^{c}Y^{I\mathbf{ij}}\ .
\end{equation}

Before we turn to higher derivative terms we wish to consider whether the 
vector 
multiplet coupling of this theory can be generalized from that presented in 
\cite{Nishino:2000cz,Nishino:2001ji}. To this end we may also add the most 
general vector multiplet coupling that is compatible with the Lagrange 
multiplier vector multiplet continuing to function as such. This reads
\begin{align}
 e^{-1}\mathcal{L}_V &= C_{IJK}\left( -\tfrac{1}{4}\rho^I F^J\cdot F^K 
+\tfrac{1}{3}\rho^I\rho^J\Box \rho^K + 
\tfrac{1}{6}\rho^I(\mathcal{D}\rho^J)\cdot(\mathcal{D}\rho^K) + \rho^I 
Y^{J\mathbf{ij}}Y^K_{\mathbf{ij}} \right.\nonumber\\
 &\left.-\tfrac{4}{3}\rho^{I}\rho^{J}\rho^{K}\left(D+\tfrac{26}{3}T^2\right) + 
4\rho^I\rho^JF^K_{\mu\nu}T^{\mu\nu}-\tfrac{e^{-1}}{24}\epsilon^{
\mu\nu\rho\sigma\lambda}A^I_\mu F^J_{\nu\rho} F^K_{\sigma\lambda}\right) \ ,
\label{generalvectors}
\end{align}
which is completely independent of the Lagrange multiplier vector multiplet.

There are two special cases where the density \eqref{generalvectors} vanishes
$\mathcal{L}_V=0$, where either $C_{IJK}=0$ or less trivially when 
$C_{IJK}=d_{(I}a_{JK)}$. To see that the density vanishes in the later the case 
note that it is formed from the combination of the vanishing composite linear 
multiplet and another set of vector multiplets, and each term in the density 
contains an element of the linear composite multiplet. One can also verify this 
by direct computation of course. Another way to see this is by considering the original cubic prepotential involving the Lagrange multiplier
vector multiplet, $\rho^\flat \mathcal{A}$. Indeed making a field redefinition of the Lagrange multiplier vector multiplet of the form
\begin{equation}
 \rho^\flat ={\rho'}^\flat + d_{I}\rho^I
\end{equation}
will not change the theory  and simply generates the vanishing term considered above.

For general $C_{IJK}$ we define
\begin{equation}
 \mathcal{C}=C_{IJK}\rho^I\rho^J\rho^K \ ,\qquad 
\mathcal{C}_I=C_{IJK}\rho^J\rho^K \ ,\qquad \mathcal{C}_{IJ}=C_{IJK}\rho^K \ ,
\end{equation}
and the density \eqref{generalvectors} becomes 
\begin{align}
 e^{-1}\mathcal{L}_V &= - 
\tfrac{1}{4}\left(\mathcal{C}_{IJ}-\tfrac{\mathcal{C}}{3}\mathcal{A}^{-1}a_{IJ}
-2\mathcal{A}^{-1}\mathcal{A}_I\mathcal{C}_J  
+\tfrac{4\mathcal{C}}{3}\mathcal{A}^{-2}\mathcal{A}_I\mathcal{A}
_J\right)F^I\cdot F^J \nonumber\\
 &+\left(\mathcal{C}_{IJ} 
-\tfrac{\mathcal{C}}{3}\mathcal{A}^{-1}a_{IJ}\right)\underline{Y}^{I\mathbf{ij}}
\underline{Y}_{\mathbf{ij}}^J 
 -\tfrac{1}{24}C_{IJK}\epsilon^{mnpqr}A^I_{m}F^J_{np}F^K_{qr}
 \nonumber\\ 
 &-\tfrac{1}{2}\left(\mathcal{C}_{IJ} 
-2\mathcal{A}^{-1}\mathcal{A}_I\mathcal{C}_J 
-\tfrac{\mathcal{C}}{3}\mathcal{A}^{-1}a_{IJ} 
+\tfrac{4\mathcal{C}}{3}\mathcal{A}^{-2}\mathcal{A}_I\mathcal{A}_J\right) 
(d\rho^I)\cdot (d\rho^J)  \nonumber\\
 &-\tfrac{1}{12}\left(\mathcal{A}^{-1}\mathcal{C}_I - 
\tfrac{2\mathcal{C}}{3}\mathcal{A}^{-2}\mathcal{A}_I\right)\epsilon^{mnpqr}F^{I}
_{mn}H_{pqr}\ .
\end{align}

Note however that this density contains terms not present in the original 
formulation, and as such this represents a generalization of the vector 
multiplet couplings, and furthermore the dilatonic couplings break the symmetry 
\eqref{shiftsymmetry}. Also note that the two Ricci scalar contributions to 
this 
density coming from the superconformal d'Alembertion have cancelled. 
Applying the transformations \eqref{scalartrans}, \eqref{gauginotransinverse} 
we 
obtain
\begin{align}
 e^{-1}\mathcal{L}_V &= 
 +e^{\sigma}\left(\tilde{\mathcal{C}}_{IJ} 
+\tfrac{\tilde{\mathcal{C}}}{3}\eta_{IJ}\right)\left(L^I_\alpha L^J_\beta 
\delta^\alpha_i\delta^\beta_j\right){Y}^{i\mathbf{ij}}{Y}_{\mathbf{ij}}^j 
+\tfrac{1}{12}\left(\tilde{\mathcal{C}}_I - 
\tfrac{2\tilde{\mathcal{C}}}{3}L_I\right)\epsilon^{mnpqr}F^{I}_{mn}H_{pqr}
 \nonumber\\ 
 &- 
\tfrac{1}{4}e^{\sigma}\left(\tilde{\mathcal{C}}_{IJ}+\tfrac{\tilde{\mathcal{C}}}
{3}\eta_{IJ}-2L_I\tilde{\mathcal{C}}_J  
 +\tfrac{4\tilde{\mathcal{C}}}{3}L_IL_J\right)F^I\cdot F^J   
-\tfrac{1}{24}C_{IJK}\epsilon^{mnpqr}A^I_{m}F^J_{np}F^K_{qr}
\nonumber\\
 &-\tfrac{1}{2}e^{3\sigma} 
\left(\tilde{\mathcal{C}}_{IJ}+\tfrac{\tilde{\mathcal{C}}}{3}\eta_{IJ} 
\right)L^I_\alpha L^J_\beta \delta^\alpha_i\delta^\beta_j (d\rho^i)\cdot 
(d\rho^j) \ .  \label{geninternalvectors}
\end{align}
 The explicit Chern-Simons term and the term involving both the 2- and 3-form 
field strengths $F^I$ and $H$ do not occur in the N-R formulation. If we demand 
their absence we find the condition
 \begin{equation}
C_{IJK}=(2\tilde{\mathcal{C}}L_{(I}-3\tilde{\mathcal{C}}_{(I})\eta_{JK)} \ 
,\label{chernsimonscancelation}
 \end{equation}
 which implies
 \begin{equation}  
\tilde{\mathcal{C}}_{IJ}=-\tfrac{\tilde{\mathcal{C}}}{3}\eta_{IJ}+2L_I\tilde{
\mathcal{C}}_J  
 -\tfrac{4\tilde{\mathcal{C}}}{3}L_IL_J \ , \label{chernsimonscancelation2}
 \end{equation}
but this implies that the entire density vanishes and we are left with the N-R 
supergravity coming from the linear multiplet density only.
Note that the Chern-Simons term clearly breaks the symmetry 
\eqref{shiftsymmetry}. Demanding \eqref{chernsimonscancelation} is the only way 
to restore it, apart from the exceptional case when we have only one vector 
multiplet in which case \eqref{chernsimonscancelation} is automatic, but the 
density in that case again vanishes as discussed in the previous section.

Now we turn to the case in which $\det{a}=0$. In this case we can still 
diagonalize the rank r tensor $a_{IJ}$ with a constant GL$(r,\mathbb{R})$ 
transformation. Putting a tilde on the indices in \eqref{generalvectors} and 
then splitting indices into $\tilde{I}=(I,\hat{I})$ with $I=(0,\cdots,r-1)$ and 
$\hat{I}=(r,\cdots n)$.  We will refer to the r $I$ directions as internal 
vector multiplets as they occur in the gravitational multiplet, and the 
remaining $\hat{I}$ directions as external vector multiplets. As the 
contribution to the density formed from the Lagrange multiplier vector 
multiplet 
and the composite linear multiplet \eqref{compositelinearmultiplet} vanishes 
for 
the external vector multiplets, we only have the contribution to the density 
\eqref{generalvectors}.
Substituting the expressions for the composite standard-Weyl multiplet fields 
this reads
 \begin{align}
 e^{-1}\mathcal{L}_V =& - 
\tfrac{1}{4}\left(\mathcal{C}_{IJ}-\tfrac{\mathcal{C}}{3}\mathcal{A}^{-1}a_{IJ}
-2\mathcal{A}^{-1}\mathcal{A}_I\mathcal{C}_J  
+\tfrac{4\mathcal{C}}{3}\mathcal{A}^{-2}\mathcal{A}_I\mathcal{A}
_J\right)F^I\cdot F^J \nonumber\\
 &- \tfrac{1}{4}\mathcal{C}_{\hat{I}\hat{J}}F^{\hat{I}}\cdot F^{\hat{J}} 
-\tfrac{1}{4}\left(2C_{I\hat{J}}-2\mathcal{A}^{-1}\mathcal{A}_{I}\mathcal{C}_{
\hat{J}}  \right)F^I\cdot F^{\hat{J}} \nonumber\\
 &+\left(\mathcal{C}_{IJ} 
-\tfrac{\mathcal{C}}{3}\mathcal{A}^{-1}a_{IJ}\right)\underline{Y}^{I\mathbf{ij}}
\underline{Y}_{\mathbf{ij}}^J  
+2\mathcal{C}_{I\hat{J}}\underline{Y}^{I\mathbf{ij}}{Y}_{\mathbf{ij}}^{\hat{J}} 
+\mathcal{C}_{\hat{I}\hat{J}}{Y}^{\hat{I}\mathbf{ij}}{Y}_{\mathbf{ij}}^{\hat{J}}
 \nonumber\\ 
 &-\tfrac{1}{2}\left(\mathcal{C}_{IJ} 
-2\mathcal{A}^{-1}\mathcal{A}_I\mathcal{C}_J 
-\tfrac{\mathcal{C}}{3}\mathcal{A}^{-1}a_{IJ} 
 +\tfrac{4\mathcal{C}}{3}\mathcal{A}^{-2}\mathcal{A}_I\mathcal{A}_J\right) 
(d\rho^I)\cdot (d\rho^J)  \nonumber\\
  &-\tfrac{1}{2}\left(2\mathcal{C}_{I\hat{J}} 
-2\mathcal{A}^{-1}\mathcal{A}_I\mathcal{C}_{\hat{J}} \right) (d\rho^I)\cdot 
(d\rho^{\hat{J}})  
  -\tfrac{1}{2}\mathcal{C}_{\hat{I}\hat{J}}(d\rho^{\hat{I}})\cdot 
(d\rho^{\hat{J}}) \nonumber\\
 &-\tfrac{1}{12}\left(\mathcal{A}^{-1}\mathcal{C}_I - 
\tfrac{2\mathcal{C}}{3}\mathcal{A}^{-2}\mathcal{A}_I\right)\epsilon^{mnpqr}F^{I}
_{mn}H_{pqr}
 -\tfrac{1}{12}\mathcal{A}^{-1}\mathcal{C}_{\hat{I}} 
\epsilon^{mnpqr}F^{\hat{I}}_{mn}H_{pqr}\nonumber\\
& 
-\tfrac{1}{24}C_{\tilde{I}\tilde{J}\tilde{K}}\epsilon^{mnpqr}A^{\tilde{I}}_{m}F^
{\tilde{J}}_{np}F^{\tilde{K}}_{qr} \ ,\label{mostgeneralvectors}
\end{align}
where
\begin{equation} 
\mathcal{C}=C_{\tilde{I}\tilde{J}\tilde{K}}\rho^{\tilde{I}}\rho^{\tilde{J}}\rho^
{\tilde{K}} \ ,\qquad 
\mathcal{C}_{\tilde{I}}=C_{\tilde{I}\tilde{J}\tilde{K}}\rho^{\tilde{J}}\rho^{
\tilde{K}} \ ,\qquad 
\mathcal{C}_{\tilde{I}\tilde{J}}=C_{\tilde{I}\tilde{J}\tilde{K}}\rho^{\tilde{K}} 
\ .
\end{equation}
 As discussed above the explicit Chern-Simons term breaks the symmetry 
\eqref{shiftsymmetry}, so if we wish to maintain it extended to the external 
vector multiplets we need that the last two lines of the above density cancel 
up 
to a surface term. In this case we immediately obtain
\begin{equation}
C_{\hat{I}\hat{J}\hat{K}}=C_{\hat{I}\hat{J}K}=0 \ ,\quad 
C_{IJK}=3\mathcal{A}^{-1}\mathcal{C}_{(I}\eta_{JK)}-2\mathcal{CA}^{-2}\mathcal{A
}_{(I}\eta_{JK)} \ ,\quad 
C_{\hat{I}JK}=\mathcal{A}^{-1}\mathcal{C}_{\hat{I}}\eta_{JK}\ , \label{vandenistyext}
\end{equation}
but again we find that in this case the density vanishes as these imply that
\begin{equation} 
\mathcal{C}_{IJ}=\tfrac{\mathcal{C}}{3}\mathcal{A}^{-1}\eta_{IJ}+2\mathcal{A}^{
-1}\mathcal{A}_I\mathcal{C}_J  
-\tfrac{4\mathcal{C}}{3}\mathcal{A}^{-2}\mathcal{A}_I\mathcal{A}_J \ .
\end{equation}
Again the vanishing of the denisty in this case can be seen from a redefinition of the Lagrange multiplier of the form
\begin{equation}
\rho^{\flat}={\rho'}^{\flat} + d_{\tilde{I}}\rho^{\tilde{I}} \ ,
\end{equation}
which generates the terms
\begin{equation}
 C_{IJK}=d_{(I}a_{JK)} \qquad C_{\hat{I}JK}=\tfrac{1}{3}d_{\hat{I}}a_{JK} \label{trivterms}
\end{equation}
which are equivalent to \eqref{vandenistyext} for some constants $d_{\tilde{I}}$.
 The density \eqref{mostgeneralvectors} is the most general vector multiplet 
coupling we can add, and we have shown that it generically breaks the symmetry 
\eqref{shiftsymmetry}. Indeed considering the original prepotential $\rho^\flat\mathcal{A} + \mathcal{C}$
which exhibits the symmetry
\begin{equation}
 \rho^\flat \rightarrow e^{-2c} \ , \qquad \rho^{I} \rightarrow e^{c}\rho^{I}  \ , \qquad \rho^{\hat{I}} \rightarrow e^{c}\rho^{\hat{I}} \ ,
\end{equation}
only in the case $\mathcal{C}=0$, up to the terms \eqref{trivterms} which can be generated by the redefinition of the Lagrange multiplier.

If however we require only that the internal vector multiplets have the 
symmetry 
\eqref{shiftsymmetry} whilst the external vector multiplets are inert under 
this 
transformation, it is clear that we may add couplings between external vector 
multiplets $\hat{I}$ whilst preserving \eqref{shiftsymmetry}, i.e we take 
$C_{IJK}=3\mathcal{A}^{-1}\mathcal{C}_{(I}\eta_{JK)}-2\mathcal{CA}^{-2}\mathcal{
A}_{(I}\eta_{JK)}$ and  
$C_{\hat{I}JK}=\mathcal{A}^{-1}\mathcal{C}_{\hat{I}}\eta_{JK}$ but now we allow 
$C_{\hat{I}\hat{J}\hat{K}}$ to be arbitrary, \footnote{We cannot have 
$C_{I\hat{J}\hat{K}}$ different from zero and maintain the symmetry as the 
corresponding Chern Simons term explicitly breaks it and there is no candidate 
cancellation term coming from the $*(F\wedge H)$ terms.} so that we maintain 
the 
symmetry
 \begin{equation}
  \sigma \rightarrow \sigma + c \qquad A^I_\mu \rightarrow e^{c}A_\mu^I \qquad 
B_{\mu\nu}\rightarrow e^{2c}B_{\mu\nu} \qquad A^{\hat{I}}_\mu \rightarrow 
A_\mu^{\hat{I}}\ , \label{shiftexternalinert}
 \end{equation}

 The density in this case reads
 \begin{equation}
 e^{-1}\mathcal{L}_V = - 
\tfrac{1}{4}\mathcal{C}_{\hat{I}\hat{J}}F^{\hat{I}}\cdot F^{\hat{J}}  
+\mathcal{C}_{\hat{I}\hat{J}}{Y}^{\hat{I}\mathbf{ij}}{Y}_{\mathbf{ij}}^{\hat{J}} 
  -\tfrac{1}{2}\mathcal{C}_{\hat{I}\hat{J}}(d\rho^{\hat{I}})\cdot 
(d\rho^{\hat{J}}) 
-\tfrac{1}{24}C_{\hat{I}\hat{J}\hat{K}}\epsilon^{mnpqr}A^{\hat{I}}_{m}F^{\hat{J}
}_{np}F^{\hat{K}}_{qr}\ . \label{externalvectorshiftinv}
\end{equation}
Note that we must therefore not transform the external scalars with our 
coordinate transformation \eqref{scalartrans}, so the supersymmetry 
transformations of the external vector multiplets are given by 
\eqref{off-shellvectorsusy}.
If we allow for different scaling behaviour of the external multiplets, we may 
construct densities which respect the symmetry
\begin{equation}
 \sigma \rightarrow \sigma + c \qquad A^I_\mu \rightarrow e^{c}A_\mu^I \qquad 
B_{\mu\nu}\rightarrow e^{2c}B_{\mu\nu} \qquad A^{\hat{I}}_\mu \rightarrow 
e^{-kc}A_\mu^{\hat{I}}\ , \label{generalizedshift}
\end{equation}
by transforming the external scalars such that 
\begin{equation}
 {\rho'}^{\hat{I}}= e^{k\sigma}\rho^{\hat{I}}\ ,
\end{equation}
and the gauginos by
\begin{equation}
 {\lambda'}^{\hat{I}}=ke^{k\sigma}\lambda^0 + e^{k\sigma}\lambda^{\hat{I}}
\end{equation}
in the following cases. We have discussed the case $k=0$ above, which 
corresponds to allowing us to take $C_{\hat{I}\hat{J}\hat{K}}$ non-zero. It is 
clear we can never take $C_{IJK}$ different from its expression above. In the 
case $k=\tfrac{1}{2}$ we may take $C_{I\hat{J}\hat{K}}\neq0$ but then we need 
$C_{\hat{I}\hat{J}\hat{K}}=0$ and $C_{IJ\hat{K}}$ must be equal to its 
expression above. Finally in the case $k=1$ we may allow $C_{\hat{I}JK}$ to 
differ from its expression above, but need 
$C_{\hat{I}\hat{J}\hat{K}}=C_{\hat{I}\hat{J}K}=0$. This can also be seen easily by inspection of the origional cubic prepotential.

The case of one internal multiplet is exceptional as we shall now discuss.
Recall that the vector density formed from the internal vector multiplet 
vanishes identically. Indeed it is also the case that a density formed from two 
internal multiplets and arbitrarily many external multiplets must vanish. This 
means that we may take arbitrary $C_{000}, C_{00\hat{I}}$, however terms 
involving these quantities will not appear in the action, and will therefore 
not 
break the symmetry \eqref{shiftsymmetry}. Indeed we may read off the most 
general contribution to the density from \eqref{mostgeneralvectors}. 
\begin{align}
 \lefteqn{e^{-1}\mathcal{L}_V =}&\label{oneinternalgenevectors}\nonumber\\
 &- 
\tfrac{1}{4}\left(e^{-\sigma}\mathcal{D}+e^{-2\sigma}\hat{\mathcal{C}}
\right)(F^0)^2 
 + 
\tfrac{1}{2}(\hat{\mathcal{C}}_{\hat{I}\hat{J}}+e^{-\sigma}\mathcal{D}_{\hat{I}
\hat{J}})F^{\hat{I}}\cdot F^{\hat{J}} +\tfrac{1}{2}\left(\mathcal{D}_{\hat{I}} 
+e^{-\sigma}\hat{\mathcal{C}}_{\hat{I}}  \right)F^0\cdot F^{\hat{I}} \nonumber\\
 &-\tfrac{1}{2}\left(\hat{\mathcal{C}} +e^{\sigma}\mathcal{D}\right) (d\sigma)^2  
  +\left(\hat{\mathcal{C}}_{\hat{I}} +e^{\sigma}\mathcal{D}_{\hat{I}} \right) 
(d\sigma)\cdot (d\rho^{\hat{I}})  
  -\tfrac{1}{2}(\hat{\mathcal{C}}_{\hat{I}\hat{J}} 
+e^{\sigma}D_{\hat{I}\hat{J}})(d\rho^{\hat{I}})\cdot (d\rho^{\hat{J}}) 
\nonumber\\
 &-\tfrac{1}{12}\left(e^{-2\sigma}\mathcal{D} - 
e^{-3\sigma}\tfrac{2}{3}\hat{\mathcal{C}}\right)\epsilon^{mnpqr}F^{0}_{mn}H_{pqr
} 
+\tfrac{1}{12}(e^{-2\sigma}\mathcal{C}_{\hat{I}}+2e^{-\sigma}\mathcal{D}_{\hat{I
}})\epsilon^{mnpqr}F^{\hat{I}}_{mn}H_{pqr}\nonumber\\
& 
-\tfrac{1}{24}C_{\hat{I}\hat{J}\hat{K}}\epsilon^{mnpqr}A^{\hat{I}}_{m}F^{\hat{J}
}_{np}F^{\hat{K}}_{qr} 
-\tfrac{1}{8}D_{\hat{I}\hat{J}}\epsilon^{mnpqr}A^{0}_{m}F^{\hat{I}}_{np}F^{\hat{
J}}_{qr}+(\hat{\mathcal{C}}_{\hat{I}\hat{J}}+e^{\sigma}D_{\hat{I}\hat{J}}){Y}^{
\hat{I}\mathbf{ij}}{Y}_{\mathbf{ij}}^{\hat{J}} 
  \ ,
\end{align}
where we defined 
\begin{equation} 
\hat{\mathcal{C}}=C_{\hat{I}\hat{J}\hat{K}}\rho^{\hat{I}}\rho^{\hat{J}}\rho^{
\hat{K}} \ ,\qquad 
\hat{\mathcal{C}}_{\hat{I}}=C_{\hat{I}\hat{J}\hat{K}}\rho^{\hat{J}}\rho^{\hat{K}
} \ ,\qquad  
\hat{\mathcal{C}}_{\hat{I}\hat{J}}=C_{\hat{I}\hat{J}\hat{K}}\rho^{\hat{K}}\ ,
\end{equation}
and
\begin{equation}
 \mathcal{D}=C_{0\hat{I}\hat{J}}\rho^{\hat{I}}\rho^{\hat{J}}\ ,\qquad 
\mathcal{D}_{\hat{I}}=C_{0\hat{I}\hat{J}}\rho^{\hat{J}} \ ,\qquad 
D_{\hat{I}\hat{J}}=C_{0\hat{I}\hat{J}}\ .
\end{equation}
Similarly to the above cases we may preserve the symmetry 
\eqref{shiftexternalinert} only if $D_{\hat{I}\hat{J}}=0$, but the theory 
exhibits a symmetry of the form \eqref{generalizedshift} after a suitable 
scalar 
and gaugino redefinition when taking only one of $D_{\hat{I}\hat{J}}$ or 
$C_{\hat{I}\hat{J}\hat{K}}$ non-vanishing. 

To summarize if we demand that the symmetry \eqref{shiftsymmetry} is extended 
to 
the external vector multiplets we may only add vector multiplet couplings of 
the 
form
\begin{equation}  
C_{IJK}=3\mathcal{A}^{-1}\mathcal{C}_{(I}\eta_{JK)}-2\mathcal{CA}^{-2}\mathcal{A
}_{(I}\eta_{JK)} \ ,\qquad 
C_{\hat{I}JK}=\mathcal{A}^{-1}\mathcal{C}_{\hat{I}}\eta_{JK}\ , 
\label{symmetrypresevedchernsimonscoeffs}
\end{equation}
with all other components zero, but the density \eqref{mostgeneralvectors} 
vanishes, and the scalar manifold is simply $SO(1,1) \times SO(1,n)/SO(n)$. On 
the other hand if we demand that the external vector multiplets are inert under 
this transformation \eqref{shiftexternalinert}, then we must take the 
expressions \eqref{symmetrypresevedchernsimonscoeffs} with 
$C_{\hat{I}\hat{J}K}=0$, but with arbitrary $C_{\hat{I}\hat{J}\hat{K}}$ and
\eqref{externalvectorshiftinv} is the corresponding density which allows for 
the 
preservation of the symmetry \eqref{shiftsymmetry}. The scalar manifold is then 
a product of $SO(1,1) \times SO(1,n)/SO(n)\times \mathcal{M}$, with 
$\mathcal{M}$ some $m=n-r$ dimensional manifold, which seems only to be 
restricted by demanding the absence of ghosts in the theory. Also an explicit 
Chern-Simons term appears.  On the other hand, if we relax the assumption that 
our theory should preserve the symmetry \eqref{shiftsymmetry} then we may add 
the general vector multiplet couplings and obtain the density 
\eqref{mostgeneralvectors}. In this case the entire scalar manifold is 
dependent 
on the form of $C_{\tilde{I}\tilde{J}\tilde{K}}$. In particular a Lagrange 
multiplier forcing a restriction of the scalar manifold, for example the very 
special geometry condition, is absent.
If we view the theory as being defined by the 
$C_{\tilde{I}\tilde{J}\tilde{K}}$ from compactification then the symmetry 
\eqref{shiftsymmetry} or even \eqref{generalizedshift} is generically broken.

\section{Higher derivative densities.}\label{Higher}
In this section we shall describe how to simply generalize the known Ricci 
squared \cite{Ozkan:2013nwa} and Weyl squared \cite{Hanaki:2006pj} invariants 
to 
an arbitrary number of internal and external vector multiplets. In 
\cite{Coomans:2012cf} an off-shell superconformal Riemann squared invariant was 
derived in the $r=1$ dilaton-Weyl multiplet that we used here to construct the 
pure N-R supergravity, but we leave the generalization of the Riemann squared 
invariant for future work.\footnote{Deriving this invariant is equivalent to 
deriving the Riemann squared invariant in the standard-Weyl multiplet, which 
has 
yet to be given in components, but was recently analysed in superspace in \cite{Butter:2014xxa}.}

\subsection{Ricci squared invariant.}

In \cite{Ozkan:2013nwa} a Ricci squared invariant coupled to vector multiplets 
in the r=1 dilaton-Weyl multiplet was constructed in a particular basis of the 
superconformal fields. This basis is equivalent to a reversible gauge fixing of 
the theory by breaking the SU(2) down to U(1), and breaking the local dilatonic 
symmetry and special supersymmetry. We shall give the details of the 
construction without going to this basis, by using the construction of the 
Ricci 
squared invariant in the standard-Weyl multiplet, which was also given in 
\cite{Ozkan:2013nwa}. The essential observation is that the Ricci scalar 
appears 
in the composite expression for the field $Y^{\mathbf{ij}}$ in terms of a 
linear 
multiplet, and that this is not cancelled by the contribution coming from the 
expression for $\underline{D}$ when moving to the general dilaton-Weyl 
multiplet. Thus in the standard-Weyl multiplet we may form the Ricci squared 
invariant by considering a composite linear multiplet, which is formed from two 
copies of a composite vector multiplet, each of which is formed from our 
compensating linear multiplet. Schematically the density is
\begin{equation}
e^{-1} \mathcal{L}= \mathbf{V}^{\tilde{I}}\cdot 
\mathbf{L}(\mathbf{V}^{\#},\mathbf{V}^{\#})
\end{equation}
where $\mathbf{V}^{\#}=\mathbf{V}(\mathbf{L}_0)$.
Clearly as the density \eqref{generalvectors} was formed from composing the 
linear multiplet from two sets of vector multiplets, we may construct a density 
from \eqref{generalvectors} by setting $C_{\tilde{I}\#\#}=e_{\tilde{I}}$, where 
the vector multiplet $\mathbf{V}^{\#}$ is composite and is formed from our 
compensating linear multiplet. After the gauge fixing 
\eqref{conformalgaugefixing} and setting $L=1$, the bosonic parts of the vector 
multiplet composed of our compensating linear multiplet, which we obtain from 
gauge fixing \eqref{compositevectoroflinear}, are simply
\begin{eqnarray}
\underline{\rho}^{\#} &=& 2 N \,, \nonumber\\
\underline{Y}^{\#}_{\mathbf{ij}} &=&\tfrac{1}{\sqrt2}\delta_{\mathbf{ij}} 
\left( 
-\tfrac{3}{8}R  - N^2  - P^2 
 + \tfrac{8}{3}  T^2+ 4 D -{V'}_a^{\mathbf{kl}}{V'}^a_{\mathbf{kl}}\right) 
\nonumber\\
 &+& 2P^a{V'}_{a \mathbf{ij}}- \sqrt2 \nabla^a 
{V'}_{a\mathbf{(i}}^{\mathbf{m}}{\delta_{\mathbf{j)m}}}    ,\nonumber\\
{F}^{\#}_{\mu\nu} &=& 4 {\partial}_{[\mu} P_{\nu ]} + 
2\sqrt2\partial_{[\mu}{V}_{\nu]}   \,.
\label{compositevectoroflineargaugefixed}
\end{eqnarray}
where we have split $V^{\mathbf{ij}}$ into its traceful and traceless parts as 
in \eqref{tracesplitV}.

We obtain the density
\begin{eqnarray}
 \lefteqn{e^{-1}\mathcal{L}_{R^2}=}&& \nonumber\\
&& 
\mathcal{E}\left(\tfrac{3}{8}R-4(\underline{D}+\tfrac{26}{3}\underline{T}
^2)+32\underline{T}^2+N^2+P^2+{V'}^2\right)^2  
-16\mathcal{E}N^2(\underline{D}+\tfrac{26}{3}\underline{T}^2)\nonumber\\
 &&+ 
2\mathcal{E}(\sqrt{2}P^a{V'}_a^{\mathbf{ij}}+\nabla^a{V'}_a^{\mathbf{ij}})(\sqrt
{2}P^a{V'}_{a\mathbf{ij}}+\nabla^a{V'}_{a\mathbf{ij}})+16\mathcal{E}N(\sqrt{2}
dV 
+2dP)\cdot \underline{T}\nonumber\\
 &&-\tfrac{1}{2}\mathcal{E}(dV)^2 -\sqrt{2}\mathcal{E} (dV)\cdot(dP) 
-\mathcal{E}(dP)^2 -2\mathcal{E}(dN)^2 
-4Ne_{\tilde{I}}(d\rho^{\tilde{I}})\cdot(dN)\nonumber\\
 &&-\sqrt{2}e_{\tilde{I}}N(F^{\tilde{I}}\cdot dV) 
-2e_{\tilde{I}}N(F^{\tilde{I}}\cdot dP)  +16e_{\tilde{I}}N^2(F^{\tilde{I}}\cdot 
\underline{T}) -\tfrac{1}{3}\mathcal{E}N^2R \nonumber \\
&& 
-2\sqrt{2} e_{\tilde{I}}\underline{Y}^{\tilde{I}\mathbf{ij}}\delta_{\mathbf{ij}}
\left(\tfrac{3}{8}RN-4\underline{D}N-\tfrac{8}{3}N\underline{T}^2+N^3+NP^2+N{V'}
^2\right)\nonumber\\
&& 
-e_{\tilde{I}}\epsilon^{\mu\nu\rho\sigma\tau}A_\mu^{\tilde{I}}\partial_{\nu}
V_\rho \partial_{\sigma}V_\tau
 -2\sqrt2 
e_{\tilde{I}}\epsilon^{\mu\nu\rho\sigma\tau}A_\mu^{\tilde{I}}\partial_{\nu}
P_\rho \partial_{\sigma}V_\tau 
-2e_{\tilde{I}}\epsilon^{\mu\nu\rho\sigma\tau}A_\mu^{\tilde{I}}\partial_{\nu}
P_\rho \partial_{\sigma}P_\tau \nonumber\\  
&&+8e_{\tilde{I}}\underline{Y}^{\tilde{I}}_{\mathbf{ij}}{V'}_a^{\mathbf{ij}}
NP^a 
-4\sqrt{2}e_{\tilde{I}}\underline{Y}^{\tilde{I}}_{\mathbf{ij}}N\nabla^a{V'}_a^{
\mathbf{m(i}}{\delta\phantom{'}}^{\mathbf{j)}}_{\mathbf{m}} \ ,
\end{eqnarray}
where $\mathcal{E}=e_{\tilde{I}}\rho^{\tilde{I}}$. Substituting the expressions 
for the composite standard-Weyl fields from \eqref{vanishinglmconditions} we 
obtain a supersymmetric Ricci squared invariant coupled to internal and 
external 
vector multiplets, whose leading term is $\tfrac{1}{4}\mathcal{E}R^2$. If we 
apply the map \eqref{scalartrans} to the internal multiplets we may add this to 
the two derivative actions derived in the previous section and the leading term 
becomes $e_{I}e^{\sigma}L^IR^2+e_{\hat{I}}\rho^{\hat{I}}R^2$, so the symmetry 
\eqref{shiftexternalinert} is maintained only in the case that we couple 
exclusively to external multiplets, i.e. $e_I=0$. If we take the point of view 
that this correction is perturbative, and since at leading order the fields 
$V_\mu^{\mathbf{ij}},Y^{\tilde{I}\mathbf{ij}}, N, P$ vanish the relevant 
contribution is  
\begin{equation}\mathcal{L}_{R^2}=\mathcal{E}(\tfrac{3}{8}R-4(\underline{D}
+\tfrac{26}{3}\underline{T}^2)+32\underline{T}^2)^2 + \cdots \ .
\end{equation}

\subsection{Weyl squared invariant.}
In \cite{Hanaki:2006pj} a supersymmetric invariant including a Weyl tensor 
squared term was constructed in the standard-Weyl multiplet and coupled to 
Abelian vector multiplets. This is given in the conventions we use in 
\cite{Ozkan:2013nwa}, which we will repeat below. We will consider the same 
construction as before, namely that we have a Lagrange multiplier vector 
multiplet coupled only to the other vectors in such a way as to implement the 
vanishing of the composite linear multiplet, providing expressions for the 
standard-Weyl fields $\underline{D}, \underline{T}_{\mu\nu}$ and 
$\underline{\chi}^{\mathbf{i}}$. In particular we will not couple the Lagrange 
multiplier vector multiplet to the higher derivative terms, and so do not 
induce 
higher derivative expressions in the definitions of these fields. 
The contribution to the bosonic action of the Weyl-squared term is given in 
\cite{Ozkan:2013nwa} and reads
\begin{eqnarray}
\lefteqn{e^{-1} \mathcal{L}_{C^2+\tfrac{1}{6} R^2} =}&& \nonumber\\
&&\beta_{\tilde{I}} \Big( \tfrac{1}{8}\rho^{\tilde{I}} {C}^{\mu\nu\rho\sigma} 
{C}_{\mu\nu\rho\sigma}+ \tfrac{64}{3} \rho^{\tilde{I}} \underline{D}^2 + 
\tfrac{1024}{9} \rho^{\tilde{I}} \underline{T}^2 \underline{D} - \tfrac{32}{3} 
\underline{D} \, \underline{T}_{\mu\nu} F^{\mu\nu\,{\tilde{I}}}   \nonumber\\
&&  - \tfrac{16}{3} \rho^{\tilde{I}} {C}_{\mu\nu\rho\sigma} 
\underline{T}^{\mu\nu} \, \underline{T}^{\rho\sigma} + 2{C}_{\mu\nu\rho\sigma} 
\underline{T}^{\mu\nu} F^{\rho\sigma\,{\tilde{I}}} + \tfrac{1}{16} 
\epsilon^{\mu\nu\rho\sigma\lambda}A_\mu^{\tilde{I}} {C}_{\nu\rho\tau\delta} 
{C}_{\sigma\lambda}{}^{\tau\delta}    \nonumber\\
&& -\tfrac{1}{12} \epsilon^{\mu\nu\rho\sigma\lambda} A_\mu^{\tilde{I}} 
{V}_{\nu\rho}{}^{ij} {V}_{\sigma\lambda\, ij} +  \tfrac{16}{3} 
Y^{\tilde{I}}_{ij} {V}_{\mu\nu}{}^ {ij} \underline{T}^{\mu\nu} - \tfrac{1}{3} 
\rho^{\tilde{I}} {V}_{\mu\nu}{}^{ij}{V}^{\mu\nu}{}_{ij} \nonumber\\
&& +\tfrac{64}{3} \rho^{\tilde{I}}  \nabla_\nu \underline{T}_{\mu\rho} 
\nabla^\mu \underline{T}^{\nu\rho} - \tfrac{128}{3} \rho^{\tilde{I}} 
\underline{T}_{\mu\nu} \nabla^\nu \nabla_\rho \underline{T}^{\mu\rho} - 
\tfrac{256}{9} \rho^{\tilde{I}} R^{\nu\rho} \underline{T}_{\mu\nu} 
\underline{T}^\mu{}_\rho \nonumber\\
&& + \tfrac{32}{9} \rho^{\tilde{I}} R \underline{T}^2 - \tfrac{64}{3}  
\rho^{\tilde{I}} \nabla_\mu \underline{T}_{\nu\rho} \nabla^\mu 
\underline{T}^{\nu\rho} + 1024 \rho^{\tilde{I}} \, 
\underline{T}_{\mu\nu}\underline{T}^{\nu\rho}\underline{T}_{\rho\sigma}
\underline{T}^{\sigma\mu}- \tfrac{2816}{27} \rho^{\tilde{I}}  
(\underline{T}^2)^2   \nonumber\\
&&- \tfrac{64}{9} {\underline{T}_{\mu\nu}} F^{\mu\nu\,{\tilde{I}}} 
\underline{T}^2 - \tfrac{256}{3} \underline{T}_{\mu\rho} 
\underline{T}^{\rho\lambda} \underline{T}_{\nu\lambda} F^{\mu\nu\,{\tilde{I}}}  
- \tfrac{32}{3}   \epsilon_{\mu\nu\rho\sigma\lambda}  \underline{T}^{\rho\tau} 
\nabla_\tau \underline{T}^{\sigma\lambda} F^{\mu\nu\,{\tilde{I}}}  \nonumber\\
&& - 16   \epsilon_{\mu\nu\rho\sigma\lambda} \underline{T}^\rho{}_\tau 
\nabla^\sigma \underline{T}^{\lambda\tau} F^{\mu\nu\,{\tilde{I}}}  - 
\tfrac{128}{3} \rho^{\tilde{I}} \epsilon_{\mu\nu\rho\sigma\lambda} 
\underline{T}^{\mu\nu} \underline{T}^{\rho\sigma} \nabla_\tau 
\underline{T}^{\lambda\tau}\Big) \ ,
\end{eqnarray}
where $\beta_{\tilde{I}}$ are constants, 
$V^{\mathbf{ij}}_{\mu\nu}=2\partial_{[\mu}V^{\mathbf{ij}}_{\nu]}-2V^{\mathbf{ik}
}_{[\mu}{V_{\nu]\mathbf{k}}}^{\mathbf{j}}$ and $C_{\mu\nu\rho\sigma}$ is the 
Weyl tensor. Note that the $D^2$ term contains a factor of the Ricci scalar 
squared, which is why we have labelled the invariant $C^2+\tfrac{1}{6}R^2$. 
This 
fact is what allows one to combine it with the Riemann squared invariant to 
form 
the Gauss-Bonnet combination \cite{Ozkan:2013uk} in the $r=1$ dilaton-Weyl 
multiplet, which is the only case that at present the Riemann squared invariant 
is known.   
Inserting the expressions for the composite fields $\underline{T}$, 
$\underline{D}$ and $\underline{Y}^{0\mathbf{ij}}$ given in 
\eqref{compositesuperconformalfields3} we obtain a supersymmetric invariant for 
arbitrary numbers of internal and external multiplets. We may then make the 
transformations \eqref{scalartrans}, \eqref{gauginotrans} in order to identify 
the dilaton. We note that the symmetry \eqref{shiftexternalinert} is broken 
unless we couple exclusively to external multiplets, i.e. $\beta_{I}=0$ and 
that 
only the third line of this invariant may be neglected in a perturbative 
treatment, due to the vanishing of the fields $V^{\mathbf{ij}}_\mu$ and 
$Y^{\tilde{I}\mathbf{ij}}$ at the two derivative level.

\section{Conclusions.}\label{Conclusions}
In this work we described in detail the construction of the $\mathcal{N}=2$ 
$d=5$ supergravity of Nishino and Rajpoot \cite{Nishino:2000cz,Nishino:2001ji} 
from the superconformal formulation 
\cite{Bergshoeff:2001hc,Fujita:2001kv,Kugo:2000hn,Kugo:2000af}. The 
construction 
of the minimal N-R model proceeded straightforwardly. In the case of the N-R 
model coupled to vector multiplets we paid particular attention to the 
identification of the dilaton amongst the scalars, and the resulting scalar 
manifolds. We found that in order for the supersymmetry transformations to be 
non-singular we must require that the homogeneous quadratic 
$\mathcal{A}=a_{IJ}\rho^I\rho^J$ must never vanish. Making the coordinate 
transformation \eqref{scalartrans} we then found it easy to identify the scalar 
manifold in the case that the only contribution from the vector multiplet 
coupling came from a quadratic coupling between them which in turn is coupled 
to 
a Lagrange multiplier vector multiplet, which gave rise to the original N-R 
formulation. It is well know that the general (two derivative) vector multiplet 
coupling is defined by a symmetric tensor $C_{IJK}$ which can be viewed as the 
triple intersection of a Calabi-Yau manifold in the compactification of 
M-theory 
\cite{Cadavid:1995bk}. From this point of view, the coupling that results in 
the 
N-R formulation is schematically
\begin{equation}
 C_{IJK}\mathbf{V}^I\cdot L(\mathbf{V}^J,\mathbf{V}^K)=\mathbf{V_{\flat}}\cdot 
L(a_{IJ}\mathbf{V}^I\mathbf{V}^J)
\end{equation}
where $a_{IJ}$ has Lorentzian signature and may be diagonalized so that in the 
new basis
\begin{equation}{a'}_{IJ}=\eta_{IJ}=\text{diag}(-1,1\cdots,1) \ .\end{equation}
As a shorthand for this we will use the notation
\begin{equation}
C_{\flat IJ}= a_{IJ} 
\end{equation}
indicating that only this component is non-zero. This can be plugged into the 
vector multiplet density \eqref{generalvectors}, and we found that the scalar 
manifold is $SO(1,1) \times SO(1,n)/SO(n)$ as described in 
\cite{Nishino:2000cz,Nishino:2001ji}.

We generalized the vector multiplet matter coupling available in the 
literature, 
but this came at the price of breaking the global scaling symmetry of the 
action 
that is present in the N-R formulation. We always consider densities that 
preserve the function of $\mathbf{V_\flat}$ as Lagrange multipliers. In 
particular first we took 
\begin{equation}
C_{\flat IJ}= a_{IJ}\ , \qquad {C'}_{IJK}\ , \label{amaxrank}
\end{equation}
non-zero and derived the density \eqref{geninternalvectors}. This generically 
breaks the shift symmetry \eqref{shiftsymmetry}, and only respects it when the 
${C'}_{IJK}$ contribution to the density vanishes, the conditions for which are 
given in \eqref{chernsimonscancelation}. We called the vector multiplets 
$\mathbf{V}^I$ in the above internal vector multiplets, as they appear in the 
gravitational multiplet. We can extend the coupling to external vector 
multiplets which do not appear in the gravitational multiplet by considering
\begin{equation}
C_{\flat IJ}= a_{IJ}, \qquad {C'}_{\tilde{I}\tilde{J}\tilde{K}}. 
\label{mostgencoupling}
\end{equation}
where $\tilde{I}=(I,\hat{I})$ and in particular does not include the $\flat$ 
direction. The form of the coefficients \eqref{amaxrank} arises from a 
compactification of the low energy limit of M-theory on a Calabi-Yau which is a 
K3 fibration \cite{Antoniadis:1995vz} where it is assumed that the rank of $a$ 
is maximal. Taking \eqref{mostgencoupling} results in the most general vector 
multiplet coupling that allows for the $\mathbf{V}_\flat$ to function as a 
Lagrange multiplier, and we gave the density in \eqref{mostgeneralvectors}. Not 
surprisingly this density generically breaks the symmetry 
\eqref{shiftsymmetry}, 
but we found that if we allow the external vector multiplets to be inert under 
these transformations we could preserve the symmetry \eqref{shiftexternalinert} 
in the particular case that we take the vector density 
\eqref{externalvectorshiftinv}, so that the scalar manifold is now a product 
$SO(1,1)\times SO(1,n)/SO(n)\times\mathcal{M}$.
We then turned to higher derivative corrections and generalized the known Ricci 
squared and Weyl squared densities to include more than one internal multiplet. 
Again these break the symmetry \eqref{shiftsymmetry}, but if we take them to be 
coupled to only external multiplets we may maintain the symmetry 
\eqref{shiftexternalinert}.

It would be interesting to explicitly consider the appropriate 
compactifications 
of the heterotic theory on suitable five manifolds and to understand better the 
relation of that theory to the off-shell theory presented here, and the duality 
to M-theory on a Calabi-Yau 3-fold. In \cite{Antoniadis:1995vz} such a 
computation was carried out using the very special geometry condition 
$\mathcal{C}=C_{\tilde{I}\tilde{J}\tilde{K}}\rho^{\tilde{I}}\rho^{\tilde{J}}
\rho^{\tilde{K}}=1$
to produce a Lagangian for the effective heterotic theory by removing one of the 
scalars from the action in the case of two internal vector multiplets, which 
is equivalent to fixing the Lagrange multiplier scalar $\rho^{\flat}$ using the 
$D$ equation of motion in the off-shell formulation, at least at the two 
derivative level.
In the heterotic 
superstring picture the presence of the additional vector couplings $C_{IJK}$ 
were related to 1-loop corrections, whilst the original N-R formulation is the 
tree level contribution. In the off-shell formulation in the standard-Weyl 
multiplet the very special geometry condition arises at the two derivative 
level 
by integrating out a Lagrange multiplier, the standard-Weyl field $D$. After 
the 
dualization we have no such constraint in the vector multiplet sector, as it 
can 
be solved using the Lagrange multiplier vector multiplet. These two approaches 
are equivalent at the two derivative level, but it seems that we ought to 
include the higher derivative corrections to the very special geometry 
constraint, or in our picture to include couplings between the higher 
derivative 
terms and the Lagrange multiplier vector multiplet, introducing higher 
derivative terms in the expression for the composite standard-Weyl fields.  For 
the case of only one internal vector multiplet the heterotic result implies the 
absence of a one loop term corresponding to the vanishing vector density of 
section \ref{Pure}, which was straightforward to show in our set-up.
It would be interesting to see how our external vector multiplets 
fit into this picture, and particularly how the higher derivative corrections 
in 
the standard- and dilaton-Weyl multiplets may be related by the 
heterotic/M-theory duality.

It would be highly desirable to derive a Riemann squared or Ricci tensor 
squared 
supersymmetric invariant in the standard-Weyl multiplet in order to construct 
arbitrary quadratic curvature supergravities. This would be of interest when 
considering higher order string theory corrections, but also within the 
framework of supersymmetric Lovelock theory or Chern-Simons supergravity 
\cite{Banados:1996hi}, although this has been investigated in a rather 
different 
approach to that we have taken here. For generic higher order theories the 
auxiliary fields of the off-shell formulation become dynamical, and in order to 
avoid this one must take a perturbative approach to integrating out these 
fields, as done in \cite{Hanaki:2006pj,Castro:2008ne}. In \cite{Ozkan:2013uk} 
it 
was shown that for the supersymmetrization of the Gauss-Bonnet term, in the 
background of a dilaton-Weyl multiplet containing only one internal vector 
multiplet, that the kinetic terms for the auxiliary fields exactly cancel, 
meaning that they can be integrated out exactly. It would be interesting to see 
if this also happens in the background of the standard-Weyl multiplet, and to 
understand the compactifications of string and M-theory to this theory. 
Interestingly the coefficients of the Chern-Simons terms which along with 
supersymmetry specify the vector multiplet couplings completely, at both the 
two 
and four derivative level, have been investigated recently in 
\cite{Bonetti:2013cza,Grimm:2014soa,Grimm:2014aha} from a 6D and M-theoretic 
perspective. Whilst this article was in preparation the interesting article 
\cite{Butter:2014xxa} appeared which addresses many of these issues from a 
superspace perspective.

We may also straightforwardly add on-shell hypermultiplet couplings to this 
theory which is desirable due to the presence of the universal hyper-multiplet 
in compactifications. This was recently discussed in \cite{Baggio:2014hua} in 
addition to higher derivative couplings. In the superconformal tensor calculus 
it is not known how to put general hypermultiplets off-shell, however in the 
superspace formulation this has been discussed in 
\cite{Kuzenko:2006mv,Kuzenko:2007cj,Kuzenko:2007hu,Kuzenko:2008wr,
Kuzenko:2013rna,Kuzenko:2014eqa} and whilst this article was in preparation the 
interesting paper \cite{Butter:2014xua} also appeared. Since the field $N$ 
appears in the linear multiplet sector which, on coupling to additional tensor 
multiplets, may provide a factor in the scalar manifold closer to the very 
special geometry of the standard formulation it would be interesting to include 
general linear multiplet couplings. It would be also be particularly 
interesting 
to gauge the models presented here, using the methods of 
\cite{Coomans:2012cf}, in particular for applications to four 
dimensional field theories via the AdS/CFT correspondence. It should also be possible to extend the internal 
gauging 
procedure of that work from gauging the internal U(1) gauge field of the 
dilaton-Weyl multiplet to gauging the full SU(2) R-symmetry using these methods 
to produce a Weyl multiplet with an internal Yang-Mills multiplet and find a 
suitable gauge fixing of the superconformal fields.

\appendix
\section{Generalized dilaton-Weyl superconformal 
multiplets.}\label{gendilatonweyl}
A general dilaton-Weyl multiplet\footnote{For $m\geq 1$.} is made up of the 
vielbien $e_\mu^a$, gravitino $\psi_\mu^{\mathbf{i}}$, $m$ gauge fields 
$A^I_{\mu}$, a two-form gauge field $B_{\mu\nu}$, $m$ scalars $\rho^I$, $m$ 
gauginos $\psi^{I\mathbf{i}}$, an auxiliary SU(2) triplet of vectors 
$V^{\mathbf{ij}}_\mu$ with $V^{\mathbf{ij}}_\mu=V^{\mathbf{ji}}_{\mu}$, $(m-1)$ 
SU(2) triplets of scalars, $Y^{i\mathbf{ij}}$ and a gauge field for local 
dilatations $b_{\mu}$. Using vector multiplet indices $I=(0,i)$ these transform 
under supersymmetry with parameter $\epsilon^{\mathbf{i}}$ and special 
supersymmetry with parameter $\eta^{\mathbf{i}}$ as

\begin{align}
 \delta e_\mu^a &=\tfrac{1}{2}\bar{\epsilon}\gamma^a\psi_\mu \ , \nonumber\\
 \delta \psi_\mu^{\mathbf{i}} &= (\nabla_\mu 
+\tfrac{1}{2}b_\mu)\epsilon^{\mathbf{i}} 
-V_{\mu}^{\mathbf{ij}}\epsilon_{\mathbf{j}} 
+i\gamma_{mn}\underline{T}^{mn}\gamma_\mu\epsilon^{\mathbf{i}} - 
i\gamma_\mu\eta^{\mathbf{i}} \ , \nonumber\\
 \delta V^{\mathbf{ij}}_\mu &= -\tfrac{3i}{2}\bar{\epsilon}^{\mathbf{(i}} 
\underline{\phi}^{\mathbf{j)}}_\mu +4\bar{\epsilon}^{\mathbf{(i}}\gamma_\mu 
\underline{\chi}^{\mathbf{j)}} + 
i\bar{\epsilon}^{\mathbf{(i}}\gamma_{mn}\underline{T}^{mn}\psi^{\mathbf{j)}}
_\mu 
+\tfrac{3i}{2}\bar{\eta}^{\mathbf{(i}}\psi^{\mathbf{j)}}_\mu  \ , \nonumber\\
 \delta b_\mu &= \tfrac{i}{2}\bar{\epsilon}\underline{\phi}{}_\mu - 
2\bar{\epsilon}\gamma_\mu\underline{\chi} + \tfrac{i}{2}\bar{\eta}\psi_\mu  \ ,  
\nonumber\\
 \delta A^I_\mu &= -\tfrac{i}{2}\rho^I\bar{\epsilon}\psi_\mu + 
\tfrac{1}{2}\bar{\epsilon}\gamma_\mu\lambda^I  \ , \nonumber\\
 \delta B_{\mu\nu} &= 
-\tfrac{1}{2}\mathcal{A}\bar{\epsilon}\gamma_{[\mu}\psi_{\nu]} - 
\tfrac{i}{2}\mathcal{A}_I \bar{\epsilon}\gamma_{\mu\nu}\lambda^I -\eta_{IJ} 
A^I_{[\mu}\delta(\epsilon)A^J_{\nu]}  \ , \nonumber\\
 \delta \lambda^{I\mathbf{i}} &= 
-\tfrac{1}{4}\gamma_{mn}\hat{F}^{Imn}\epsilon^{\mathbf{i}} - 
\tfrac{i}{2}\gamma^a(\mathcal{D}_a\rho^I)\epsilon^{\mathbf{i}} + 
\rho^I\gamma_{mn}\underline{T}^{mn}\epsilon^{\mathbf{i}} - 
\underline{Y}^{I\mathbf{ij}}\epsilon_{\mathbf{j}} + \rho^I\eta^{\mathbf{i}}  \ 
, 
\nonumber\\
\delta Y^{i\mathbf{ij}} &= 
-\tfrac{1}{2}\bar{\epsilon}^{\mathbf{(i}}\gamma^m\mathcal{D}_m\lambda^{\mathbf{
j)}i} + 
\tfrac{i}{2}\bar{\epsilon}^{\mathbf{(i}}\gamma_{mn}\underline{T}^{|mn|}\lambda^{
\mathbf{j)}i} - 
4i\rho^i\bar{\epsilon}^{\mathbf{(i}}\underline{\chi}^{\mathbf{j)}} + 
\tfrac{i}{2}\bar{\eta}^{\mathbf{(i}}\lambda^{\mathbf{j)}i} \ ,\nonumber\\
   \delta \rho^I &= \tfrac{i}{2}\bar{\epsilon}\lambda^I  \ , 
\label{conformalsusytransEDW}
\end{align}
where the spin covariant derivative is defined by
\begin{equation}
 \nabla_\mu\epsilon^\mathbf{i}=(\partial_\mu 
+\tfrac{1}{4}{{\omega}_{\mu}}^{mn}\gamma_{mn})\epsilon^{\mathbf{i}} \ ,
\end{equation}
where
\begin{equation}
{{\omega}}_\mu^{mn}= 2e^{\nu[m}\partial_{[\mu} e^{n]}_{\nu]} - 
e^{\nu[m}e^{n]}\sigma e_{\mu p}\partial_\nu e_\sigma^p + 2e_{\mu}^{[m}b^{n]} 
-\tfrac{1}{2}\bar{\psi}^{[n}\gamma^{m]}\psi_\mu 
-\tfrac{1}{4}\bar{\psi}^n\gamma_\mu\psi^m \ ,
\end{equation}
and we have underlined composite fields, expressions for which are given by
\begin{align}
 \underline{\phi}^\mathbf{i}_\mu 
&=\tfrac{i}{3}\gamma^m\underline{\hat{R'}}_{\mu 
m}^{\mathbf{i}}(Q)-\tfrac{i}{24}\gamma_\mu \gamma^{mn} 
{\hat{\underline{R}'}}_{mn}^{\mathbf{i}}(Q)\ , \nonumber\\
\underline{\hat{R'}}_{\mu \nu}^{\mathbf{i}}(Q) &= 
2\nabla_{[\mu}\psi^{\mathbf{i}}_{\nu]} + b_{[\mu}\psi^{\mathbf{i}}_{\nu]} - 
2V^{\mathbf{ij}}_{[\mu}\psi_{\nu]\mathbf{j}} + 
2i\gamma_{mn}\underline{T}^{mn}\gamma_{[\mu}\psi^{\mathbf{i}}_{\nu]}\ 
,\nonumber\\
\underline{\hat{R}}_{\mu \nu}^{\mathbf{i}}(Q) &=\underline{\hat{R'}}_{\mu 
\nu}^{\mathbf{i}}(Q) -2i\gamma_{[\mu}\underline{\phi}_{\nu]}^{\mathbf{i}} \ , 
\label{compositesuperconformalfields2}
\end{align}
as in the standard-Weyl multiplet but now we also have
\begin{align}
\underline{T}^{mn} &=   
-\tfrac{1}{8\mathcal{A}}\left(\tfrac{1}{6}\epsilon^{mnpqr}\hat{H}_{pqr}  
-\mathcal{A}_I\hat{F}^{Imn}- 
\eta_{IJ}\tfrac{i}{4}\bar{\lambda}^I\gamma^{mn}\lambda^J\right)\ ,\nonumber\\
\underline{\chi}^{\mathbf{i}} &= \eta_{IJ}\mathcal{A}^{-1}\left( 
\tfrac{i}{8}\rho^I\gamma^{m}\mathcal{D}_m\lambda^{J\mathbf{i}} 
+\tfrac{i}{16}\gamma^{m}(\mathcal{D}_m\rho^I)\lambda^{J\mathbf{i}} 
\right.\nonumber\\
&\left.- \tfrac{1}{32}\gamma_{mn}\hat{F}^{Imn}\lambda^{J\mathbf{i}} + 
\tfrac{1}{4}\rho^I\gamma_{mn}\underline{T}^{mn}\lambda^{J\mathbf{i}} 
-\tfrac{1}{8}\underline{Y}^I_{\mathbf{ij}}\lambda^{J\mathbf{j}}\right)\ , 
\nonumber\\
\underline{D} &= -\tfrac{26}{3}\underline{T}^2 + \eta_{IJ}\mathcal{A}^{-1} 
\left( \tfrac{1}{4}\rho^{I}\Box\rho^J 
+\tfrac{1}{8}(\mathcal{D}\rho^I)(\mathcal{D}\rho^J) - 
\tfrac{1}{16}\hat{F}_{mn}^I\hat{F}^{Jmn} 
-\tfrac{1}{8}\bar{\lambda}^I\gamma^m\mathcal{D}_m\lambda^J \right.\nonumber\\
&\left.+ \tfrac{1}{4}\underline{Y}^I_{\mathbf{ij}}\underline{Y}^{J\mathbf{ij}} 
-4i\rho^{I}\lambda^{J}\underline{\chi}  +\left(2\rho^I\hat{F}^J_{mn}  
+\tfrac{i}{4}\bar{\lambda}^I\gamma_{mn}\lambda^J \right)\underline{T}^{mn} 
\right)  \ ,\nonumber\\ 
\underline{Y}^{\mathbf{ij}0}&=(\rho^0)^{-1}\mathcal{A}_{i}{Y}^{\mathbf{ij}i}
-\tfrac{i}{4}\eta_{IJ}(\rho^0)^{-1}\bar{\lambda}^{I\mathbf{i}}\lambda^{J\mathbf{
j}} \ ,\label{compositesuperconformalfields3}
 \end{align}
where
\begin{equation}
\mathcal{A}=\eta_{IJ}{\rho^I\rho^J}\ , \qquad \mathcal{A}_I=\eta_{IJ}\rho^J\ 
,\qquad \eta_{IJ}=\text{diag}(-,+,\cdots,+) \ .
\end{equation}

As discussed at length in the main body of the text the dilaton of the N-R 
formulation is to be identitified as $\sigma=\tfrac{1}{2}\ln{(-\mathcal{A})}$ 
and we need $\mathcal{A}\neq0$ for the expressions 
\eqref{compositesuperconformalfields3} to be non-singular. 
\section{Explicit field redefinition.}
\label{explictredef}
Here we give the explicit field redefinitions needed to arrive at the N-R 
formulation in the notation of \cite{Nishino:2000cz,Nishino:2001ji}. Starting 
from the on-shell theory with Lagrangian \eqref{on-shellbetterconventions}, 
which is invariant under supersymmetry transformations 
\eqref{susyon-shellbetterconventions} we need to make the following field 
redefinitions
 \begin{align}
  \epsilon^{\mathbf{i}} &= -\sqrt2{\epsilon'}^\mathbf{i}\ , \qquad 
\psi^{\mathbf{i}}_\mu=-\sqrt2 {\psi'}^{\mathbf{i}}_\mu \ , \qquad &A^I_\mu 
=\sqrt2 {A'}^I_\mu \ , \qquad V_a^\alpha =\tfrac{1}{\sqrt{2}}{V'}_a^\alpha \ , 
\nonumber\\
  \chi^{\mathbf{i}} &= -\tfrac{\sqrt2}{\sqrt3}{\chi'}^{\mathbf{i}} \ , \qquad 
\lambda^{a\mathbf{i}} = -\sqrt{2} {\lambda'}^{a\mathbf{i}}\ ,  &  \qquad 
V^a_\alpha =\sqrt{2}{V'}^a_\alpha \ ,
 \end{align}
 and redefine 
 \begin{equation}
  L^I_\alpha=\sqrt2{L'}^I_\alpha \ ,\qquad 
L^\alpha_I=\frac{1}{\sqrt2}{L'}^\alpha_I \ .
 \end{equation}

The definition of the three form field strength, which be now call $G$, has 
therefore changed to 
 \begin{equation}                                                                            
{G'}_{\mu\nu\rho}=3\partial_{[\mu}B_{\nu\rho]} - 
3\eta_{IJ}{A'}^I_{[\mu}{F'}^J_{\nu\rho]} \ ,
\end{equation}
where ${F'^I}=d{A'}^I=\sqrt2 F^I$
and the metric is rescaled to
\begin{equation}
  {g'}_{\alpha\beta}=\tfrac{1}{2}\left(\delta_{\alpha\beta} - 
\tfrac{1}{(L^0)^2}\delta_{\alpha\gamma}\delta_{\beta\delta}\varphi^{\gamma}
\varphi^{\delta}\right)=\tfrac{1}{2}g_{\alpha\beta} \ .
\end{equation}

Note that $L^I_A$ and the SO(n) connection 1-form $A$ remains unchanged, 
however 
the parameter $\xi$ has now become $\xi=-\tfrac{1}{\sqrt2}$, due to the 
appearance of the vielbein in \eqref{cosetdiffidentities}. It is not difficult 
to see that on can further rescale the vielbein $V^a_\alpha=k {V'}^a_\alpha$, 
$V^a_\alpha=\tfrac{1}{k} {V'}^a_\alpha$ leaving $L^I_A$ fixed and redefining 
the 
scalars $\varphi=\tfrac{1}{k}{\varphi}'$ whilst leaving all other fields fixed. 
The Lagrangian and supersymmetry transformations are invariant under this map, 
however the explicit expressions for the $L^I_A$ in terms of the scalars will 
change. This is equivalent to scaling the spacelike directions in our 
coordinate 
transformations \eqref{scalartrans} and \eqref{gauginotrans} and so the choice 
of the parameter $\xi$ is, in this way, arbitrary. For each fixed value of the 
dilaton the physical scalar manifold with metric $g_{\alpha\beta}$ is a cone. 
The full scalar manifold including the dilaton is clearly the solid cone, and 
what we have described 
is a foliation by the dilaton of the full scalar manifold, whose leaves are 
hyperboloids of equal constant Ricci curvature. A different choice of the value 
of the Ricci scalar is then just an alternative foliation.
The bosonic part of the action is now given by
\begin{align}
\mathcal{L}=&-\frac{1}{4}R  - 
\frac{1}{4}e^{-2\sigma}({L_I}^aL_{Ja}+L_IL_J)F_{\mu\nu}^I F^{\mu\nu 
J}-\frac{1}{12}e^{-4\sigma}G^2 \nonumber\\
&-\frac{1}{2}g_{\alpha\beta}(d\varphi^{\alpha})\cdot(d\varphi^\beta)- 
\frac{3}{4}(d \sigma)^2
\end{align}
and its fermionic completion up the quadratic order in the fermions is given in 
\cite{Nishino:2000cz,Nishino:2001ji}.
The supersymmetry transformations, up to quadratic order in fermions read:
\begin{align}
\delta {e_\mu}^m &= \bar{\epsilon} \gamma^m \psi_{\mu} \ , \qquad 
\delta\sigma=\tfrac{i}{\sqrt{3}}\bar{\epsilon} \chi \ , \nonumber \\
\delta {\psi_\mu}^{\mathbf{i}} &= D_{\mu}\epsilon^{\mathbf{i}} + 
\tfrac{i}{6\sqrt{2}}e^{-\sigma}\Big( 
{\gamma_{\mu}}^{\rho\sigma}-4{\delta_\mu}^\rho \gamma^{\sigma}\Big) 
\epsilon^{\mathbf{i}} L_I {F_{\rho\sigma}}^I+\tfrac{1}{18}e^{-2\sigma}\Big( 
{\gamma_{\mu}}^{\rho\sigma\tau}-\tfrac{3}{2}{\delta_\mu}^\rho 
\gamma^{\sigma\tau}\Big)\epsilon^{\mathbf{i}} G_{\rho\sigma\tau} \nonumber \ ,\\
\delta {A_{\mu}}^I &=-\tfrac{i}{\sqrt{2}}e^{\sigma}L^I\bar{\epsilon}\psi_{\mu } 
+ \tfrac{1}{\sqrt{6}}e^{\sigma}L^I\bar{\epsilon}\gamma_{\mu}\chi + 
\tfrac{1}{\sqrt{2}}e^{\sigma}\bar{\epsilon}\gamma_{\mu}\lambda^a{L_a}^I \ , 
\nonumber \\
\delta B_{\mu\nu} &=e^{2\sigma}\bar{\epsilon}\gamma_{[\mu}\psi_{\nu]} + 
\tfrac{i}{\sqrt{3}}e^{2\sigma}\bar{\epsilon}\gamma_{\mu\nu}\chi-2L_{IJ}{A_{[
\mu\vert}}^I\delta_Q {A_{\vert \nu ]}}^J \ , \nonumber \\
\delta \chi^{\mathbf{i}} &= -\tfrac{1}{2\sqrt{6}} e^{-\sigma} 
\gamma^{\mu\nu}\epsilon^{\mathbf{i}} L_I 
{F_{\mu\nu}}^I+\tfrac{i}{6\sqrt{3}}e^{-2\sigma}\gamma^{\mu\nu\rho}\epsilon^{
\mathbf{i}} G_{\mu\nu\rho} 
-\tfrac{\sqrt{3}i}{2}\gamma^{\mu}\epsilon^{\mathbf{i}} \partial_{\mu}\sigma \ , 
\nonumber \\
\delta \varphi^\alpha &=\tfrac{i}{\sqrt{2}}{V_a}^{\alpha}\bar{\epsilon} 
{\lambda^a}\ , \qquad \delta_Q 
\lambda^{a\mathbf{i}}=-\tfrac{1}{2\sqrt{2}}e^{-\sigma}\gamma^{\mu\nu}\epsilon^{
\mathbf{i}} {L_I}^a {F_{\mu\nu}}^I 
-\tfrac{i}{\sqrt{2}}\gamma^\mu \epsilon^{\mathbf{i}} {V_{\alpha}}^a 
\partial_{\mu}\varphi^\alpha \ .
\label{susymatterts}
\end{align}
\section{Vector multiplet composed of a linear multiplet.} \label{vlcomposition}
One can also construct the elements of vector multiplet in terms of the 
elements 
of a linear multiplet and a Weyl multiplet \cite{Coomans:2012cf, Ozkan:2013uk, 
Ozkan:2013nwa}.
Here we just list the bosonic parts
\begin{eqnarray}
\rho &=& 2 L^{-1} N \,, \nonumber\\
Y_{\mathbf{ij}} &=& L^{-1} \Box^C L_{\mathbf{ij}} - 
{\cal{D}}_{a}L_{\mathbf{k(i}} {\cal{D}}^{a}L_{\mathbf{j)m}} L^{\mathbf{km}} 
L^{-3} - N^2 L_{\mathbf{ij}} L^{-3} - P_{\mu}P^{\mu}L_{\mathbf{ij}} 
L^{-3}\nonumber\\
&& + \tfrac{8}{3} L^{-1} T^2 L_{\mathbf{ij}} + 4 L^{-1} D L_{\mathbf{ij}} + 2 
P_{\mu} L_{\mathbf{k(i}} {\cal{D}}^{\mu} L_{\mathbf{j)}}{}^{\mathbf{k}} L^{-3}   
,\nonumber\\
{F}_{\mu\nu} &=& 4 {\cal{D}}_{[\mu}(L^{-1} P_{\nu ]}) + 2L^{-1} 
{R}_{\mu\nu}{}^{\mathbf{ij}}(V) L_{\mathbf{ij}} - 2 L^{-3} 
L_{\mathbf{k}}^{\mathbf{l}} {\cal{D}}_{[\mu}L^{\mathbf{kp}} {\cal{D}}_{\nu ]} 
L_{\mathbf{lp}}   \,.
\label{compositevectoroflinear}
\end{eqnarray}
where the bosonic parts of the relevant covariant derivatives, d'Alembertion 
and 
the curvatures are given by
\begin{eqnarray} 
\mathcal{D}_{\mu}L^{\mathbf{ij}}&=&(\nabla_\mu-3b_\mu)L^{\mathbf{ij}}+2V^{
\phantom{\mu}\mathbf{(i}}_{\mu\phantom{\mathbf{(i}}\mathbf{k}}L^{\mathbf{j)k}}_{
\phantom{\mathbf{k}}}\nonumber\\
 \mathcal{D}_{\mu}P_{\nu}&=&(\nabla_{\mu}-4b_{\mu})P_\nu \nonumber\\
 \Box^C L^{\mathbf{ij}} &=&(\nabla^a-4b^a)\mathcal{D}_{a}L^{\mathbf{ij}} + 
2V_{a\ \mathbf{k}}^{\mathbf{(i}}\mathcal{D}^aL^{\mathbf{j)k}}_{\phantom{a\ 
\mathbf{k}}}+6L^{\mathbf{ij}}f_a^a\nonumber\\ 
R(V)_{\mu\nu}^{\mathbf{ij}}:&=&V^{\mathbf{ij}}_{\mu\nu}=2\partial_{[\mu}V^{
\mathbf{ij}}_{\nu]} - 2V_{[\mu}^{\mathbf{k(i}}V_{\nu]\mathbf{k}}^{\mathbf{j})}
\end{eqnarray}
and for closure of the algebra the constraint $\mathcal{D}^{a}P_a=0$ is needed, 
and $f_a^a$ is given in \eqref{compositesuperconformalfields}.
\bibliographystyle{JHEP}
\bibliography{offshellnr}

\end{document}